\documentclass{article}

\usepackage{PRIMEarxiv}

\usepackage[utf8]{inputenc} 
\usepackage[T1]{fontenc}    
\usepackage{hyperref}       
\usepackage{url}            
\usepackage{booktabs}       
\usepackage{amssymb}
\usepackage{amsthm}
\usepackage{amsmath}
\usepackage{amsfonts}       
\usepackage{nicefrac}       
\usepackage{microtype}      
\usepackage{lipsum}
\usepackage{fancyhdr}       
\usepackage{color}
\usepackage{subfigure}
\usepackage{graphicx}       
\graphicspath{{media/}}     

\usepackage{tikz}
\usepackage{algorithmic}
\usepackage{algorithm,float}

\newtheorem{assumption}{Assumption}
\newtheorem{problem}{Problem}
\newtheorem{theorem}{Theorem}
\newtheorem{remark}{Remark}
\newtheorem{lemma}{Lemma}
\newtheorem{proposition}{Proposition}

\newtheorem{corollary}{Corollary}

\newtheorem{property}{Property}
  
\title{Simultaneous Recursive Identification of Parameters and Switching Manifolds Identification of Discrete-Time Switched Linear Systems
}

\author{
  Zengjie Zhang \\
  Eindhoven University of Technology \\
  Eindhoven, Netherlands\\
  \texttt{z.zhang3@tue.nl} \\
   \And
  Yingwei Du \\
  Sany Heavy Industry Co., Ltd \\
  Suzhou, China\\
  \texttt{yinwei.du@hotmail.com} \\
  \AND
  Tong Liu \\
  Technical University of Munich \\
  Munich, Germany \\
  \texttt{tong.liu@tum.de} \\
  \And
  Fangzhou Liu \\
  Harbin Institute of Technology \\
  Harbin, China \\
  \texttt{fangzhou.liu@hit.edu.cn} \\
  \And
  Martin Buss \\
  Technical University of Munich \\
  Munich, Germany \\
  \texttt{mb@tum.de} \\
}

\begin{document}
\maketitle

\begin{abstract}
A novel procedure for the online identification of a class of discrete-time switched linear systems, which simultaneously estimates the parameters and switching manifolds of the systems, is proposed in this paper. Firstly, to estimate the parameters of the subsystems, a discrete-time concurrent learning-based recursive parameter estimator is designed to guarantee the exponential convergence of the estimation errors to zero. Secondly, as an assistant procedure of the identification framework, an online switching detection method is proposed by making use of the history stacks produced by the concurrent learning estimators. Thirdly, techniques of incremental support vector machine are applied to develop the recursive algorithm to estimate the system switching manifolds, with its stability proven by a Lynapunov-based method. At the end of the paper, the stability and precision of the proposed identification methods are confirmed by the numerical simulation of a 2-order switched linear system. Compared to the traditional offline identification methods, the proposed online identification framework possesses superior efficiency with respect to large amounts of data, while the limitations and outlook of this framework are also discussed within the conclusion.
\end{abstract}


\section{Introduction}
The hybrid dynamic systems have attracted many attentions over the past decade due to their wide applications on depicting complex physical dynamics and engineering problems~\cite{lai2010identification,lv2018simultaneous,bao2018macroscopic,su2018sliding}. The switched linear systems, which are mathematically described by Piece-Wise Linear (PWL) models, are a typical type of hybrid systems that consist of several subsystems defined on different regions in the system space and activated by different switching conditions~\cite{heemels2001equivalence, sontag1981nonlinear}. The regions of the subsystems are separated by switching manifolds, or sub-spaces that define the switching conditions of the system. As a elementary issue of PWL systems, the identification problem is always a challenging work, since it contains several coupled sub-issues, such as the estimation of subsystem parameters and identification of switching manifolds. Surveys on formulations and solutions of these problems can be found in~\cite{paoletti2007identification, garulli2012survey}. 

A large amount of previous work has contributed to solving the identification of switched linear systems. Apart from the primitive optimization-based~\cite{wen2007identification,taguchi2009identification}, algebraic~\cite{vidal2003algebraic, vidal2008recursive} or clustering-based~\cite{baptista2011split, nakada2005identification, ferrari2003clustering, gegundez2008identification} methods proposed by earlier work, the most recent work focuses more on complex data regression models, such as regularized regression~\cite{ohlsson2013identification}, improved least squares~\cite{shah2014parameter,breschi2016piecewise} and the progressive learning machine~\cite{yang2012bidirectional,yang2015progressive}. However, the majority of these methods are merely capable of offline identification which relies on the offline data samples from previous system operations, and the computational load tremendously increases when the data scale becomes larger. In comparison, the online or recursive identification methods that identify the system simultaneously with the system sequential sampling are more efficient when handling large amounts of data, nevertheless, more challenging to implement. In recent work, only a few online parameter estimation methods for switched linear systems are proposed, such as the recursive parameter identification for switched regressive exogenous~\cite{vidal2008recursive,wang2011online} and PWL models~\cite{bako2011recursive,kersting2014online,kersting2014concurrent,kersting2017recursive}. To the best of our knowledge, there is very little work on the recursive identification of switching manifolds for switched linear systems, which impedes the direction towards a complete online identification framework for hybrid systems.

In this paper, we propose a recursive identification procedure for discrete-time switched linear systems which is capable of recursively estimating system parameters and switching manifolds simultaneously with system sequential sampling. The contributions of this work are threefold. Firstly, for the first time, a complete online identification framework for a class of hybrid systems is proposed without knowledge on system parameters nor switching manifolds. Secondly, an exponential converging parameter estimator and an online switching detection method are proposed based on the discrete-time application of concurrent learning techniques. Thirdly, a recursive switching manifold estimation method is proposed based on incremental support vector machine (SVM), and the convergence of the estimation errors is proven under proper assumptions.

The paper is organized as follows. Section \ref{sec:pre} introduces the system model and elementary assumptions investigated in this paper. The detailed results of the online identification framework is presented in section \ref{sec:rst}. In section \ref{sec:ns}, the proposed methods are evaluated by numerical simulation. Finally, section \ref{sec:cf} concludes the paper.

\section{Preliminaries}\label{sec:pre}

In this paper, we consider the discrete-time switched linear systems of the following form with $N$ subsystems
\begin{equation}\label{eq:sys}
\begin{split}
    x_{k+1} &= A(\Omega^i) x_{k} + B(\Omega^i) u_{k}, 
\end{split}
\end{equation}
where $x_k \in \Omega \subseteq \mathbb{R}^n$ and $u_k \in \mathbb{R}^m$ are respectively the state and input vector samples of the system at time instant $k$, and $\Omega$ denotes the state space. System parametric matrices $A(\Omega^i) \in \mathbb{R}^{n \times n}$ and $B(\Omega^i) \in \mathbb{R}^{n \times m}$ are dependent on the polyhedral regions $\Omega^i \subset \Omega$,
\begin{equation}
        A(\Omega^i) = A^i,~B(\Omega^i) = B^i,~\mathrm{if}~x_k \in \Omega^i,~\forall~i \in \mathcal{N},
\end{equation}
where $A^i$, $B^i$ are constant matrices and $\mathcal{N} = \{1,2,\cdots,N \}$ is the set of region numbers. Each region $\Omega^i$ is a partition of the state space $\Omega$ and surrounded by $\mu_i$ switching manifolds
\begin{equation}
    \chi^i = \left\{\chi_{1}^{i},~\chi_{2}^{i},~\cdots,~\chi_{r}^i,~\cdots,~\chi_{\mu_i}^{i} \right\},~1 \leqslant r \leqslant \mu_i,
\end{equation}
where $\chi_{r}^{i}$ denotes the $r$-th manifold covering the region $\Omega^i$. In this paper, we only consider systems as (\ref{eq:sys}) with the following properties.

\begin{property}\label{pt:pt1}
The switching manifolds $\chi_{r}^{i}$, $1 \leqslant r \leqslant \mu_i$ are linear semi-hyper-planes passing through and split by the state space origin $\Omega_o$, i.e.
\begin{equation}
\Omega_o \in \bigcap \limits^{N}_{i=1} \bigcap \limits^{\mu_i}_{j=1}  \chi^{i}_{j},
\end{equation}
and can be represented by
\begin{equation}
    \chi_{r}^{i} = \left\{ \left.x\right| x^T h_{r}^{i} = 0,~\eta^i_r(x) > 0,~x \in \Omega\right\},~i \in \mathcal{N},
\end{equation}
where $h_{r}^{i} \in \mathbb{R}^n$ is the coefficient vector of $\chi^i_r$ and $\eta^i_r(x)=0$ denotes the boarder of semi-hyper-plane $\chi_{r}^{i}$ with another switching manifold. Thus we can use the vector set $\mathcal{H}^i = \left\{h_{1}^{i},~h_{2}^{i},~\cdots,~h_{\mu_i}^{i} \right\}$ to depict $\chi^i$, and each region $\Omega^i$ can be defined as the intersections of $\mu_i$ half spaces split by these hyper-planes, i.e.
\begin{equation}
    \Omega^i = \left\{ \left. x \right| x^T h < 0, \forall h \in \mathcal{H}_i \right\},~i \in \mathcal{N}.
\end{equation}
\end{property}

\begin{property}\label{pt:pt2}
The regions $\Omega^i$ of the system (\ref{eq:sys}) in the state space $\Omega$ are completely and mutually exclusively defined, i.e.
\begin{equation}
    \Omega = \bigcup \limits^N_{i=1} \Omega^i,~\Omega^j \cap \Omega^r = \varnothing,~\forall j,r \in \mathcal{N} ,~j \neq r.
\end{equation}
\end{property}

Further more we propose the following assumption for the work in this paper.
\begin{assumption}\label{as:uasum}
    If the initial state of the system is $x_0 \in \Omega^i$, $i \in \mathcal{N}$ and $x_0 \neq \Omega_o$, then there exists at least one input sequence $\mathcal{U}^i_{\tau} = \left\{ u^i_0,u^i_1,\cdots,u^i_{\tau} \right\}$, $\tau < + \infty$, such that the generated system state sequence $\mathcal{X}^i_{\tau} = \left\{ x^i_1,x^i_2,\cdots,x^i_{\tau+1} \right\}$ satisfies the following conditions.
    
    (1). $\mathcal{X}^i_{\tau}$ traverses all $N$ regions, i.e. $\mathcal{X}^i_{\tau} \cap \Omega^j \neq \varnothing $, $\forall~j \in \mathcal{N}$;
    
    (2). Supposing that $\mathcal{X}^i_{\tau_j^1} = \{ x^i_{j_1}, x^i_{j_2}, \cdots, x^i_{j_{\tau_j}} \}  \subset \mathcal{X}^i_{\tau}$ denotes the first successive state sequence of the system staying in region $\Omega^j$, there exists a minimum dwelling time $\tau_{\min} = \min \limits_{j \in \mathcal{N}} {\tau_j}$ for all $\mathcal{X}^i_{\tau_j^1}$, $j \in \mathcal{N}$;
    
    (3). The system input $u_k$ at each instant $k$ is bounded as $\|u_k\| \leqslant \delta_u$, such that the state increment $\Delta x_k = x_{k+1} - x_k$ is also bounded as $\| \Delta x_k \| \leqslant \delta_x$;
    
    (4). There exist a positive definite matrix $P \in \mathbb{R}^{n \times n}$ and a scalar $0 < \epsilon < 1$, such that for the positive definite scalar function $V_k = x_k^T P x_k$, $V_{k+1} <  \epsilon V_k$, $\forall~k \geqslant 0$ holds.
\end{assumption}

\begin{remark}
The first two conditions in Assumption \ref{as:uasum} are also referred to as the \textit{long time dwelling} assumption which is frequently used related work on identification of hybrid systems~\cite{petreczky2011notion, baptista2011split}, which is intended to guarantee enough data samples for the identification in each region. The third condition confines the state samples to be close enough to each other to improve the density of the sample distribution, and the third condition supposes the input sequence $\mathcal{U}^i_{\tau}$ to guarantee the global exponential stability of the system in the equilibrium $\Omega_o$~\cite{zhai2001stability}. Justifications on these conditions will be discussed in Section \ref{sec:rst}.
\end{remark}

Here we only assumes that such input sequences $\mathcal{U}^i_{\tau}$ exist without presenting the specific solutions, although it is known that the adaptive control methods can be applied to obtain such control inputs~\cite{kersting2017direct}. Other methods on the input design for identifications of hybrid systems can be found in~\cite{suzuki2011input}. It is worth mentioning that the persistent excitation condition, which is also commonly used in some previous work~\cite{petreczky2011notion, kersting2017direct, di2013hybrid}, is not required in this work.

\section{Main Results}\label{sec:rst}
The goal of this paper is to propose an online identification procedure for system (\ref{eq:sys}) to recursively estimate the parameters $A^i$ and $B^i$, and the switching manifolds $\mathcal{H}^i$ for all subsystem regions $\Omega^i$, $i \in \mathcal{N}$. To achieve this, the whole identification problem is decomposed into three subproblems, namely recursive parameter estimation, online system switching detection and recursive switching manifold estimation. These subproblems are solved independently but implemented simultaneously within the online identification framework. Before presenting the results, we rewrite the system (\ref{eq:sys}) as the following regression form
\begin{equation}\label{eq:sysphi}
    x_{k+1} = \Phi^i d_k,~x_k \in \Omega^i,
\end{equation}
where $\Phi^i = \left[ A^i ~B^i \right]$ and $d_k = \left[  x_k^T ~ u_k^T \right]^T$. Here we refer to $x_{k+1}$ as the new data sample, while all $u_i$, $x_i$, $\forall~0\leqslant i \leqslant k$ as the history data samples.

\subsection{Recursive Parameter Estimation}\label{sec:rpe}
The recursive parameter estimation problem for system (\ref{eq:sysphi}) is formulated as follows.

\begin{problem}\label{pb:para}
Given the system initial state $x_0 \in \Omega^{i_0}$, $i_0 \in \mathcal{N}$, $x_0 \neq \Omega_o$, and the input and state sequences $\mathcal{U}^{i_0}_{\tau}$, $\mathcal{X}^{i_0}_{\tau}$ such that the conditions in Assumption \ref{as:uasum} hold, construct the recursive estimators $\hat{\Phi}^i_k \rightarrow \hat{\Phi}^i_{k+1}$ for all $i \in \mathcal{N}$, such that the estimation $\hat{\Phi}^i_k$ converges to the true value of $\Phi^i$ as $k$ increases.
\end{problem}

A concurrent learning based discrete-time parameter estimator is proposed for this problem by the following theorem. 

\begin{theorem} \label{th:thepara}
For a given scalar $ 0 < \sigma < 1$, if there exists a positive definite matrix $\Gamma \in \mathbb{R}^{n}$, such that $\left(\Gamma - I \right)^2 < \sigma I$ then the following recursive estimation law
\begin{equation}\label{eq:law}
    \hat{\Phi}^i_{k+1} = \left\{ \begin{array}{ll}
    \hspace{-0.1cm}     \hat{\Phi}^i_{k} - \Gamma R^i_k \left(D^i_{k} \right)^{-1}, & k \geqslant p^i, \\
    \hspace{-0.1cm}     \hat{\Phi}_k^i,& 0 \leqslant k < p^i
    \end{array} \right. 
\end{equation}
guarantees the exponential decay of estimation error $\tilde{\Phi}^i_k = \hat{\Phi}^i_k - \Phi^i$ to zero when $k \geqslant p^i$, where $R^i_k$ and $D^i_k$ are respectively referred to as the history residual stack and the history data stack of subsystem $i$ at time instant $k$ and defined as
\begin{equation}\label{eq:hisstk}
R^i_k = \sum \limits^{k}_{j=1} r^i_j d_{j-1}^T,~D^i_k = \sum \limits_{j=1}^k d_{j-1} d_{j-1}^T,~\mathrm{if}~x_j \in \Omega^i,
\end{equation}
with the residual vector $r_j^i$ defined as
\begin{equation}\label{eq:resieps}
r_j^i =  \hat{\Phi}^i_k d_{j-1} - x_j,~\mathrm{if}~x_k \in \Omega^i,
\end{equation}
and integer $p^i > 0$ is the minimum number such that $D^i_p$ is non-singular.
\end{theorem}

\begin{proof}
    Substituting (\ref{eq:sysphi}) and (\ref{eq:resieps}) to (\ref{eq:hisstk}) we have
    \begin{equation}\label{eq:phidev}
        R^i_k = \sum \limits_{j=1}^{k} r_j^i d_{j-1}^T = \tilde{\Phi}^i_k \sum \limits_{j=1}^{k} d_{j-1} d_{j-1}^T = \tilde{\Phi}^i_k D^i_{k},
    \end{equation}
    and the estimation error $\tilde{\Phi}^i_{k+1}$ reads
    \begin{equation}\label{eq:ddevphi}
        \tilde{\Phi}^i_{k+1} = \left\{ \begin{array}{ll}
             \tilde{\Phi}^i_{k} - \Gamma \tilde{\Phi}^i_{k}, & k \geqslant p^i, \\
             \tilde{\Phi}^i_{k}, & 0 \leqslant k< p^i.
        \end{array} \right.
    \end{equation}
    We define the following discrete-time Lyapunov function
    \begin{equation}
        V^i_k = \frac{1}{2} \mathrm{tr} \left( \tilde{\Phi}^{iT}_k \Gamma^{-1} \tilde{\Phi}^i_k \right),
    \end{equation}
    and calculate the increment $\Delta V^i_k = V^i_{k+1} - V^i_k$ as
    \begin{equation}
         \Delta V^i_k=\mathrm{tr} \left( \tilde{\Phi}_k^{iT} \Gamma^{-1} \Delta \tilde{\Phi}^i_k \right) + \frac{1}{2} \mathrm{tr} \left( \Delta \tilde{\Phi}_k^{iT} \Gamma^{-1} \Delta \tilde{\Phi}^i_k \right),
    \end{equation}
    where $\Delta \tilde{\Phi}^i_k = \tilde{\Phi}^i_{k+1} - \tilde{\Phi}^i_k $. By substituting (\ref{eq:ddevphi}) into the equation above we have, for $ 0 \leqslant k < p^i$, $\Delta V^i_k =0$ and for $k \geqslant p^i$, 
	\begin{equation}\label{eq:dv}
        \Delta V^i_k = - \mathrm{tr} \left( \tilde{\Phi}_k^{iT} \tilde{\Phi}^i_k \right) + \frac{1}{2} \mathrm{tr} \left(  \tilde{\Phi}_k^{iT} \Gamma  \tilde{\Phi}^i_k  \right) 
		=\mathrm{tr}\left[ \tilde{\Phi}_k^{iT} \left( \frac{1}{2} \Gamma - I \right) \tilde{\Phi}^i_k \right] .
	\end{equation}	    
	According to $\left( \Gamma - I \right)^2 < \sigma \Gamma$, it is easy to obtain
	\begin{equation}\label{eq:gi}
		\frac{1}{2} \Gamma - I < \frac{\sigma - 1}{2} \Gamma^{-1}.
	\end{equation}
    Therefore, (\ref{eq:dv}) and (\ref{eq:gi}) leads to
	\begin{equation}
		\Delta V^i_k = \mathrm{tr}\left[ \tilde{\Phi}_k^{iT} \left( \frac{1}{2} \Gamma - I \right) \tilde{\Phi}^i_k \right] < \frac{1}{2} \left(\sigma - 1 \right) \mathrm{tr} \left( \tilde{\Phi}^{iT}_k \Gamma^{-1} \tilde{\Phi}^i_k \right) < (\sigma - 1) V^i_k.
	\end{equation}	    
    Thus we have $V^i_{k+1} = \sigma^{-k} V^i_k$, which leads to
	\begin{equation}
		V_k^i \left\{ \begin{array}{ll}
			< \sigma^{-k} \left( \sigma^{-p^i} V^i_{p^i} \right), & k \geqslant p^i, \\
			= V_0 < +\infty, & 0 \leqslant k <p^i. 
		\end{array} \right.
	\end{equation}	    
    Thus the exponential convergence of the estimation $\hat{\Phi}^i_k$ is proven, and this holds $\forall i \in \mathcal{N}$.
\end{proof}

\begin{remark}\label{rk:concur}
The solution (\ref{eq:law}) presented in Theorem \ref{th:thepara} is a discrete-time application of the concurrent learning techniques, which relies on history stacks $R^i_k$ and $D^i_k$ storing history data to correct the estimation errors.. Other applications such as continous-time concurrent learning observers can be found in~\cite{kersting2014concurrent, chowdhary2010concurrent}. Note that the estimation only starts when the history data stack $D^i_k$ collects enough linearly independent data samples and become non-singular. Considering this, the minimum dwelling time $\tau_{\mathrm{min}}$ in Assumption \ref{as:uasum} should require
\begin{equation}\label{eq:mindwl}
\tau_{\mathrm{min}} \geqslant \max \limits_i p^i,
\end{equation}
such that enough samples are collected before the system switches out from the current region $\Omega^i$. It is worth mentioning that $D^i_k$ should always be non-singular $\forall~k \geqslant p^i$, where advanced techniques such as history stack purging or erroronous data removing can be applied~\cite{kersting2015removing}.
\end{remark}

\begin{remark}
It is noticed that the history stacks $R^i_k$ and $D^i_k$ in (\ref{eq:hisstk}) only collect history data belonging to $\Omega^i$. Therefore, it is necessary to label each sample with their belonging regions which are not known in advance, such that the history stacks are correctly calculated. Of course this labeling process should be conducted simultaneously with system sequential samplings. In section \ref{sec:ossd}, we will show the online system switching detection method is able to label samples recursively, but it requires $D^i_k$ to be non-singular which has been discussed in Remark \ref{rk:concur}.
\end{remark}

Compared to the traditional parameter identification for switched linear systems, the proposed concurrent learning based estimators (\ref{eq:law}) has several advantages. Firstly, the estimation is recursively performed to fit the online application. Secondly, the persistent excitation condition of the system input is not required. Thirdly, the minimum dwelling time is merely required for the first stay of the system in each region, and the condition (\ref{eq:mindwl}) is also less conservative than the traditional \textit{long time dwelling} assumptions. Finally, 
the estimation is safe from system switching, since the history stacks take turn to collect samples when the system is switched. Therefore, the concurrent learning method is quite suitable for the online identification framework.

\subsection{Online System Switching Detection}\label{sec:ossd}
In section \ref{sec:rpe}, we remarked the importance of online labeling of the samples. To achieve this, any system switching should be detected recursively, such that the samples are correctly labeled. Therefore, we formulate the following online switching detection problem for system (\ref{eq:sys}).

\begin{problem}\label{pb:ssd}
Given the system initial state $x_0 \in \Omega^i_0$, $i_0 \in \mathcal{N}$, $x_0 \neq \Omega_o$, and the input and state sequences $\mathcal{U}^{i_0}_{\tau}$, $\mathcal{X}^{i_0}_{\tau}$, such that the conditions in Assumption \ref{as:uasum} hold,  recursively label data sample $x_k$ with $n_k$, $n_k \in \mathcal{N}$, when a new data sample $x_{k+1}$ is observed, such that $x_k \in \Omega^{n_k}$ is true.
\end{problem}

Let us suppose the current sample $x_k \in \Omega^j$, $j \in \mathcal{N}$, and the state predictions for $k+1$ based on the parameter estimations $\hat{\Phi}^r_k$, $\forall~r \in \mathcal{N}$ as in (\ref{eq:law}) are $\hat{x}_{k+1}^r = \hat{\Phi}^r_k d_{k}$, and the prediction deviation reads
\begin{equation}
\hat{x}_{k+1}^r - x_{k+1} = \left( \hat{\Phi}^r_k - \Phi^j  \right) d_k = \left( \tilde{\Phi}^r_k + \Phi^r_k - \Phi^j \right) d_k,
\end{equation}
where $\tilde{\Phi}^r_k$ can be obtained from history stacks by calculating $\tilde{\Phi}_k^r = R^r_k \left(D^r_k \right)^{-1}$ according to (\ref{eq:phidev}). We thus define an auxiliary decision variable $\lambda^r_k$ for region $\Omega^r$ as
\begin{equation}\label{eq:etak}
\lambda^r_{k} = \left\| \hat{x}_{k+1}^r - x_{k+1} - R^r_k (D^r_k)^{-1} d_k \right\|,
\end{equation}
and we have
\begin{equation}\label{eq:dif}
\begin{split}
\lambda^r_{k} \left\{ \begin{array}{cc}
= 0,~\mathrm{if}~r = j, \\
\neq 0,~\mathrm{if}~r \neq j.
\end{array}   \right.
\end{split}
\end{equation}

In practice, when a new data $x_{k+1}$ is sampled, we have $n_I$ identified regions $\Omega^{I_1}$, $\Omega^{I_2}$, $\cdots$, $\Omega^{I_{n_I}}$ whose history stacks are full and parameters are estimated, and $n_U$ unidentified regions $\Omega^{U_1}$, $\Omega^{U_2}$, $\cdots$, $\Omega^{U_{n_U}}$ whose history stacks are empty or undefined and parameters are not estimated. Therefore, the decision variable in (\ref{eq:etak}) should be calculated for all the identified regions. When $\lambda^s_k \neq 0$ holds for all identified regions, an unidentified region is \textit{discovered}, and a new decision variable is generated for it. Considering this, we propose the following algorithm for Problem \ref{pb:ssd}.

\begin{algorithm}\label{ag:dtc}
\caption{Online System Switching Detection}
\begin{algorithmic}[1]
\renewcommand{\algorithmicrequire}{\textbf{Data:}}
\renewcommand{\algorithmicensure}{\textbf{Results:}}
\REQUIRE new data sample $x_{k+1}$,
\ENSURE  label $n_k$ for data $x_k$.
\STATE Calculate $\lambda^{\rho}_{k}$ according to (\ref{eq:etak}), $\forall I_1 \leqslant \rho \leqslant I_{n_i}$;
	\IF {$\lambda^{n_{k-1}}_k < \delta $}
	\STATE $n_k = n_{k-1}$;	 
 	\ELSIF {($\lambda^{I_j}_k<\delta$)}
 	\STATE $n_k = I_j$;
 	\ELSE 
 	\STATE $n_k = \rho+1$ \& $\rho=n_k$ \& Create new stacks $R^{\rho}_k$ and $D^{\rho}_k$;
 	\ENDIF
 	\IF {$n_k \neq n_{k-1}$}
 	\STATE tsslc($n_k$, $n_{k-1}$, $k$);
 	\ENDIF
\end{algorithmic} 
\end{algorithm}

In Algorithm (\ref{ag:dtc}), $\delta \in \mathbb{R}$ is a small scalar as the detection threshold. The statements 9-11 is the training sample selection process for the recursive switching manifold estimation which will be explained in section \ref{sec:rsme}. The advantage of Algorithm \ref{ag:dtc} is that the detection can be conducted even when the estimation errors $\left\| \tilde{\Phi}^i_k \right\|$ are not zero, and merely causes one step delay attributing to the history stacks, i.e. $x_{k+1}$ labels $x_k$, which is quite timely compared to existing recursive switching detection methods~\cite{lee2013line}.

\begin{remark}
The recursive parameter estimator in (\ref{eq:law}) and the online switching detection scheme as Algorithm \ref{ag:dtc} iteratively depend on each other. Specifically, when a new data $x_{k+1}$ is sampled, $\hat{\Phi}^i_k$, $\forall~i \in \mathcal{N}$ and history stacks $R^i_k$, $D^i_k$ are obtained from (\ref{eq:law}) and $x_k$ is labeled as $n_k$. Then $x_k$ is pushed into stacks $R^{n_k}_k$, $D^{n_k}_k$ and $\hat{\Phi}^{n_k}_{k+1}$ is calculated by (\ref{eq:law}). Therefore, the two independently developed method are capable to work simultaneously.
\end{remark}

Compared to the clustering based methods used for the labeling of offline samples~\cite{nakada2005identification, gegundez2008identification, baptista2011split}, the online switching detection scheme takes use of the system information revealed in the sequential sampling process, such that parameter identification and sample labeling can are achieved simultaneously. As a result, the recursive labeling scheme consumes fewer computational resources.

\subsection{Recursive Switching Manifold Estimation}\label{sec:rsme}

\subsubsection{Manifold Estimation and SVM}

In this chapter we discuss the recursive estimation of switching manifolds $\mathcal{H}^i$ for system (\ref{eq:sys}). The traditional offline identification problem has been formulated as a classification problems and solved by supervised learning based methods, such as SVM based methods~\cite{boukharouba2009identification}. It is known that the discriminant function $f:\mathbb{R}^n \rightarrow \mathbb{R}$ of a SVM classifier for a two-classification problem given training set $\mathcal{T} = \{x_{t_0},x_{t_1},\cdots,x_{t_k} \}$ is denoted as
\begin{equation}\label{eq:difgen}
    f(x_c) = \sum \limits_{i=0}^{{k}} \alpha^i l_i \kappa(x_{t_i},x_c) + b,
\end{equation}
where $x_{t_i} \in \mathbb{R}^n$, $i=0,1,\cdots,{k}$ are training samples in $\mathcal{T}$, $x_c \in \mathbb{R}^n$ is the sample to be classified, $l_i = \{-1,1\}$ are the class labels of the samples $x_{t_i}$, $\kappa:\mathbb{R}^n \times \mathbb{R}^n \rightarrow \mathbb{R}$ is the kernel function and $\alpha^i \in \mathbb{R}$ are the weights of each sample $x_{t_i}$ composing the SVM in (\ref{eq:difgen}) which are the solutions of the following quadratic optimization problem
\begin{equation}\label{eq:opt}
\begin{split}
    \min \limits_{i,j} & \left[\sum \limits_{i=0}^{k} \sum \limits_{j=0}^{k}  \alpha^i \alpha^j l_i l_j  \kappa(x_{t_i},x_{t_j})  - \sum \limits_{i=0}^{k} \alpha^i + b \sum \limits_{i=0}^{k} l_i \alpha^i  \right], \\
    \mathrm{s.t.}& \sum \limits_{i=0}^{k} l_i \alpha^i = 0~\mathrm{and}~0 \leqslant  \alpha^i \leqslant C,~\forall~x_{t_i},x_{t_j} \in \mathcal{T},
\end{split}
\end{equation}
where $C$ is the regularization parameter to punish the erroneously classified samples. Note that if the samples are linearly classifiable, the \textit{linear} kernel function is used as
\begin{equation}\label{eq:kernel}
\kappa(x_{t_i},x_{t_j}) = x_{t_i}^TQx_{t_j},~x_{t_i},x_{t_j} \in \mathcal{T},
\end{equation}
where positive definite matrix $Q \in \mathbb{R}^{n \times n}$ implements the linear transformation which will be discussed later. In this sense, the discriminant function (\ref{eq:difgen}) can be formulated as the following linear discriminant form
\begin{equation}\label{eq:ldc}
f(x_c) = w^Tx_c +b,~w = \sum \limits_{i=0}^{k} \alpha^i l_i Q x_{t_i},
\end{equation}
and the optimal solution to (\ref{eq:opt}) is geometrically equivalent to a pair of hyper-planes $g(x) = 0$, where $g(x)= \pm f(x) -1$, and we have for each data sample $x_{t_i} \in \mathcal{T}$,
\begin{equation}\label{eq:gvl}
g(x_{t_i}) = l_i f(x_{t_i}) -1 \left\{ \begin{array}{cl}
> 0, & \alpha^i = 0, \\
= 0, & 0 < \alpha^i < C, \\
< 0, & \alpha^i = C.
\end{array} \right.
\end{equation}
For brevity we use $g^i$ to represent $g(x_{t_i})$, and according its relation with $\alpha^i$ in (\ref{eq:gvl}), we define the following sample sets
\begin{equation}
\mathcal{S} = \{ x_{t_i} | 0 < \alpha^i < C \},~\mathcal{B} = \{x_{t_i} | \alpha^i = C\}, 
\mathcal{O} = \{x_{t_i} | \alpha^i = 0 \},~x_{t_i} \in \mathcal{T},
\end{equation}
where $\mathcal{S}$, $\mathcal{B}$ and $\mathcal{O}$ respectively stand for sets of support vectors, bounded support vectors and non-support vectors. 

Let us suppose $\Omega^r$ and $\Omega^j$, $r,j \in \mathcal{N}, r \neq j$ are adjacent regions in $\Omega$ and separated by switching manifolds, or semi-hyper-planes $\chi^r_{{t_k}_r}$ and $\chi^j_{{t_k}_j}$, $1 \leqslant {t_k}_r \leqslant \mu_r$, $1 \leqslant {t_k}_j \leqslant \mu_j$. Note that although $\chi^r_{{t_k}_r}$ and $\chi^j_{{t_k}_j}$ are geometrically overlapped, their parametric vectors are opposite to each other, r.e. $h^r_{{t_k}_r} = - h^j_{{t_k}_j}$. Therefore, the data samples in $\Omega^r$ and $\Omega^j$ as well as their labels obtained from Algorithm \ref{ag:rme}, can be used to train a SVM classifier. If the distribution of the samples guarantees that the trained SVM discriminant hyper-plane $f(x)=0$ is adequately close to the switching manifold $\chi^r_{{t_k}_r} $ or $ \chi^j_{{t_k}_j}$, then it can be claimed that 
\begin{equation}\label{eq:wbo}
w \approx \pm h^r_{{t_k}_r} \mp h^j_{{t_k}_j}~\mathrm{and}~b \approx 0
\end{equation}
hold. Therefore, $b = 0$ can be asserted for the discriminant model (\ref{eq:ldc}), and note that the sign in (\ref{eq:wbo}) has no influence on the precision of manifold estimation.  

The last step to adapt SVM method to the switching manifold estimation problem is to generate binary class labels $l_{k} = \{1,-1\}$ for samples $x_{t_k}$ from their region labels $n_{t_k}$ for each sample $x_{t_k} \in \mathcal{T}$ as
\begin{equation}
l_{k} = \left\{ \begin{array}{ll}
1,&x_{t_k} \in \Omega^r, \\
-1,&x_{t_k} \in \Omega^j,
\end{array} \right.
\end{equation}
and keep in mind that we have $l_{k}^2 = 1$, $\forall x_{t_k} \in \mathcal{T}$.

Based on the above statements, we are ready to formulate the following recursive switching manifolds identification problem for system (\ref{eq:sys}).

\begin{problem}\label{pb:olme}
Supposing regions $\Omega^r$ and $\Omega^j$, $r,j \in \mathcal{N}, r \neq j$ are separated by switching manifold $\chi^r_{{t_k}_r}$ or $\chi^j_{{t_k}_j}$, the training set $\mathcal{T}^{rj}_{k} = \{x_{t_0},x_{t_1},\cdots,x_{t_k} \}$ contains the sequential data samples in $\Omega^r$ or $\Omega^j$, i.e. $\mathcal{T}^{rj}_{k} \subset \Omega^r \cup \Omega^j$, and $w_{t_k}$ is the weight vector of the SVM classifier that fits the training set $\mathcal{T}^{rj}_{k}$, develop the recursive algorithm $w_{t_k} \rightarrow w_{{t_{k+1}}}$ for new sample $x_{{t_{k+1}}}$, such that $w_{{t_{k+1}}}$ fits $\mathcal{T}^{rj}_{{{k+1}}} =  \mathcal{T}^{rj}_{{k}} \cup x_{{t_{k+1}}}$ , and $w_{t_k}$ converges to the true value of $h^r_{{t_k}_r}$ or $h^j_{{t_k}_j}$ as ${k}$ increases.
\end{problem}

\subsubsection{Basic Idea of Incremental SVM}

Different from the traditional offline classifier model training, the online identification method should recursively and incrementally tune the SVM model every time a new sample $x_{t_{k+1}}$ is observed, such that the tuned model fits the newly combined training set $\mathcal{T}^{rj}_{{{k+1}}}$. For brevity we ignore superscript of the training set and represent it as $\mathcal{T}_{{{k+1}}}$. To solve Problem \ref{pb:olme}, the incremental SVM based methods~\cite{cauwenberghs2001incremental,laskov2006incremental} are applied. The main idea of incremental SVM is the following KKT balance equation that holds during the incremental tuning for the new data sample $x_{{t_{k+1}}}$,
\begin{equation}\label{eq:constr}
 \Delta g^i = \sum \limits_{i=0}^{{{k+1}}} l_i l_j \kappa(x_{t_i},x_{t_j}) \Delta  \alpha^j,~
 0 = \sum \limits_{i=0}^{{{k+1}}} l_j \Delta \alpha^j,~x_{t_i} \in \mathcal{T}_{k} \cup x_{{t_{k+1}}},
\end{equation}
where $\Delta \alpha^i$ and $\Delta g^i$ are respectively the increments of $\alpha^i$ and $g^i$ during the incremental tuning. If the elements of sample sets $\mathcal{S}$, $\mathcal{B}$ and $\mathcal{O}$ are invariant during the incremental tuning, we may as well order the elements as $\mathcal{S} = \{x_{s_1}, x_{s_2}, \cdots, x_{s_{n_s}}\}$, $\mathcal{B} = \{x_{b_1}, x_{b_2}, \cdots, x_{b_{n_b}}\}$ and $\mathcal{O} = \{x_{o_1}, x_{o_2}, \cdots, x_{o_{n_o}}\}$. Therefore, it can be obtained from (\ref{eq:gvl}) that
\begin{equation}\label{eq:zrzr}
	\Delta g^i = 0,~\forall x_i \in \mathcal{S}~\mathrm{and}~\Delta \alpha^i = 0,~\forall x_{t_i} \in \mathcal{B}\cup \mathcal{O}.
\end{equation}

For brevity and clarification, here we define the important symbols as follows: $\alpha^{\mathcal{S}}_{t_k}$ = $[\alpha_{t_k}^{s_1}~\alpha_{t_k}^{s_2}~\cdots~\alpha_{t_k}^{s_{n_s}}]^T$ and $\alpha^{{{k+1}}}_{t_k}$ respectively denote the $\alpha$ weights of the support vectors in $\mathcal{S}$ and the new sample $x_{{t_{k+1}}}$ before the incremental tuning for new sample $x_{{t_{k+1}}}$, while $\alpha^{\mathcal{S}}_{{t_{k+1}}}$ and $ \alpha^{{t_{k+1}}}_{{t_k}}$ are after the tuning, and $\Delta \alpha^{\mathcal{S}}_{t_k}$ = $\alpha^{\mathcal{S}}_{{t_{k+1}}} - \alpha^{\mathcal{S}}_{{t_k}}$, $\Delta \alpha^{{{k+1}}}_{t_k} = \alpha^{{{k+1}}}_{{t_{k+1}}} - \alpha^{{{k+1}}}_{t_k}$ are the increments. Similarly, $g^{\mathcal{B}}_{t_k}$ = $[g_{t_k}^{b_1}~g_{t_k}^{b_2}~\cdots~g_{t_k}^{b_{n_b}}]^T$, $g^{\mathcal{O}}_{t_k}$ = $[g_{t_k}^{o_1}~g_{t_k}^{o_2}~\cdots~g_{t_k}^{o_{n_o}}]^T$ , $g^{{{k+1}}}_{{t_{k}}}$are respectively the $g$ values of the samples in $\mathcal{B}$ and $\mathcal{O}$ before the incremental tuning, while $g^{\mathcal{B}}_{{t_{k+1}}}$, $g^{\mathcal{O}}_{{t_{k+1}}}$, $g^{{{k+1}}}_{{t_{k+1}}}$ are after, and $\Delta g^{\mathcal{B}}_{t_k}$ = $g^{\mathcal{B}}_{{t_{k+1}}} - g^{\mathcal{B}}_{{t_k}}$, $\Delta g^{\mathcal{O}}_{t_k}$ = $g^{\mathcal{O}}_{{t_{k+1}}} - g^{\mathcal{O}}_{{t_k}}$, $\Delta g^{{{k+1}}}_{t_k} = g^{{{k+1}}}_{{t_{k+1}}} - g^{{{k+1}}}_{t_k}$ are the increments.
It is noticed that we use superscripts to denote the owner of the $\alpha$ and $g$ values, and subscripts to denote the time when the incremental tuning is conducted. Importantly, we assert $\alpha^{{{k+1}}}_{t_k} = 0$, since $x_{{t_{k+1}}}$ is not involved in the tuning for $\mathcal{T}_{k}$.

Based on these representations, we can rewrite (\ref{eq:constr}) and (\ref{eq:zrzr}) as the following simplified form
\begin{equation}\label{eq:mat1}
\Xi_{\mathcal{S}\mathcal{S}} \Delta \alpha^{\mathcal{S}}_{t_k} = - \xi^{\mathcal{S}}_{{t_{k+1}}} \Delta \alpha^{{{k+1}}}_{{t_k}},~
l_{\mathcal{S}}^T \Delta \alpha^{\mathcal{S}}_{t_k} = -l_{k+1} \Delta \alpha^{{{k+1}}}_{t_k} 
\end{equation}
and
\begin{equation}\label{eq:mat2}
\Delta g^{\mathcal{B}}_{t_k} = \Xi_{\mathcal{B}\mathcal{S}} \Delta \alpha^{\mathcal{S}}_{t_k} + \xi^{\mathcal{B}}_{{t_{k+1}}} \Delta \alpha^{{{k+1}}}_{t_k},~
\Delta g^{\mathcal{O}}_{t_k} = \Xi_{\mathcal{O}\mathcal{S}} \Delta \alpha^{\mathcal{S}}_{t_k} + \xi^{\mathcal{O}}_{{t_{k+1}}} \Delta \alpha^{{{k+1}}}_{t_k},
\end{equation}
where $l_{\mathcal{S}} = [l_{s_1}~l_{s_2}~\cdots~l_{s_{n_s}}]^T$ is the vector of class labels of the support vectors in $\mathcal{S}$, and the vectors $\xi^{\mathcal{S}}_{{t_{k+1}}} \in \mathbb{R}^{n_s}$, $\xi^{\mathcal{B}}_{{t_{k+1}}} \in \mathbb{R}^{n_b}$, $\xi^{\mathcal{O}}_{{t_{k+1}}} \in \mathbb{R}^{n_o}$ and the matrices $\Xi_{\mathcal{S}\mathcal{S}} \in \mathbb{R}^{{n_s} \times {n_s}}$, $\Xi_{\mathcal{B}\mathcal{S}} \in \mathbb{R}^{{n_b} \times {n_s}}$, $\Xi_{\mathcal{O}\mathcal{S}} \in \mathbb{R}^{{n_o} \times {n_s}}$ are element-wisely assigned as
\begin{equation}\label{eq:vm}
\begin{split}
&\left( \xi^{\mathcal{S}}_{{t_{k+1}}} \right)_i = l_{k+1} l_{s_i} \kappa \left( x_{{t_{k+1}}},x_{s_i} \right),~
\left(\Xi_{\mathcal{SS}}\right)_{ij} = l_{s_i} l_{s_j} \kappa \left(x_{s_i},x_{s_j} \right), \\
&\left( \xi^{\mathcal{B}}_{{t_{k+1}}} \right)_i = l_{k+1} l_{b_i} \kappa \left( x_{{t_{k+1}}},x_{b_i} \right),~
\left(\Xi_{\mathcal{BS}}\right)_{ij} = l_{b_i} l_{s_j} \kappa \left(x_{b_i},x_{s_j} \right), \\
&\left( \xi^{\mathcal{O}}_{{t_{k+1}}} \right)_i = l_{k+1} l_{o_i} \kappa \left( x_{{t_{k+1}}},x_{o_i} \right),~
\left(\Xi_{\mathcal{OS}}\right)_{ij} = l_{o_i} l_{s_j} \kappa \left(x_{o_i},x_{s_j} \right).
\end{split}
\end{equation}
where $(\cdot)_i$ denotes the $i$-th element of the vector, and $(\cdot)_{ij}$ denotes the $i$-th row and $j$-th column of the matrix. It is noticed that these vectors and matrices in (\ref{eq:vm}) are only dependent on the sample labels and kernel functions, and we use their superscripts and subscripts to denote the samples that they are connected with. Further, we respectively rewrite (\ref{eq:mat1}) and (\ref{eq:mat2}) into more compact forms as follows
\begin{equation}\label{eq:comp}
\begin{split}
&\overline{\Xi}_{\mathcal{SS}}  \overline{\Delta \alpha}^{\mathcal{S}}_{t_k} = \overline{\xi}^{\mathcal{S}}_{{t_{k+1}}} \Delta \alpha^{{{k+1}}}_{t_k}, \\
&\Delta g^{\mathcal{BO}}_{{t_k}} = \Xi_{\mathcal{BOS}} \Delta \alpha^{\mathcal{S}}_{{t_k}} + \xi^{\mathcal{BO}}_{{t_{k+1}}} \Delta \alpha^{{{k+1}}}_{t_k},
\end{split}
\end{equation}
where
\begin{equation}
\begin{split}
\overline{\Xi}_{\mathcal{SS}} = \left[ \begin{array}{cc}
0 & l_{\mathcal{S}}^T \\
l_{\mathcal{S}} & \Xi_{\mathcal{SS}}
\end{array} \right],~\overline{\Delta \alpha}^{\mathcal{S}}_{t_k}= \left[ \begin{array}{c}
0 \\
\Delta \alpha^{\mathcal{S}}_{t_k} 
\end{array} \right],~\overline{\xi}^{\mathcal{S}}_{{t_{k+1}}} = \left[ \begin{array}{c}
l_{k+1} \\
\xi^{\mathcal{S}}_{{t_{k+1}}}
\end{array} \right],\\
\Delta g^{\mathcal{BO}}_{{t_k}} =
\left[ \begin{array}{c}
\Delta g^{\mathcal{B}}_{t_k} \\
\Delta g^{\mathcal{O}}_{t_k}
\end{array} \right],~\Xi_{\mathcal{BOS}} = \left[ \begin{array}{c}
\Xi_{\mathcal{BS}} \\
\Xi_{\mathcal{OS}}
\end{array} \right],~\xi^{\mathcal{BO}}_{{t_{k+1}}} =
\left[ \begin{array}{c}
\xi^{\mathcal{B}}_{{t_{k+1}}} \\
\xi^{\mathcal{O}}_{{t_{k+1}}}
\end{array} \right].
\end{split}
\end{equation}
If $\overline{\Xi}_{\mathcal{SS}}$ is non-singular, $\overline{\Delta \alpha}^{\mathcal{S}}_{t_k}$ can be calculated by
\begin{equation}\label{eq:mcp}
\overline{\Delta \alpha}^{\mathcal{S}}_{t_k} = \overline{\Xi}_{\mathcal{SS}}^{-1} \overline{\xi}^{\mathcal{S}}_{{t_{k+1}}} \Delta \alpha^{{{k+1}}}_{t_k}
\end{equation}
with given $\Delta \alpha^{{{k+1}}}_{t_k}$, and thus $\Delta \alpha^{\mathcal{S}}_{{t_k}}$ is obtained. Further more, $\Delta g^{\mathcal{B}}_{t_k}$ and $\Delta g^{\mathcal{O}}_{t_k}$ can be calculated by (\ref{eq:comp}), such that $\alpha^{\mathcal{S}}_{t_{k+1}}$, $g^{\mathcal{B}}_{t_{k+1}}$ and $g^{\mathcal{B}}_{t_{k+1}}$ are updated for $w_{k+1}$.

In practice, the sample sets may vary during the incremental tuning due to the saturation of the $\alpha$ weights of the support vectors. As a result, samples may migrate across different sets and the KKT balance equation in (\ref{eq:constr}) does not hold~\cite{cauwenberghs2001incremental}. Therefore, one incremental tuning should be conducted for several iterations. In each iteration, the saturation of sample weights should be detected, such that (\ref{eq:constr}) is recalculated. Considering this, the following algorithm for switching manifold estimation for system (\ref{eq:sys}) is presented.

\begin{algorithm}[ht]
\caption{Algorithm for Recursive Manifold Estimation}
\label{ag:rme}
\begin{algorithmic}[1]
\renewcommand{\algorithmicrequire}{\textbf{Input:}}
\renewcommand{\algorithmicensure}{\textbf{Output:}}
\REQUIRE $x_{{t_{k+1}}}$
\ENSURE  $w_{{t_{k+1}}}$
 	\STATE $\alpha^{{{k+1}}}_{t_k} = 0$;
  \WHILE {$g^{{{k+1}}}_{t_k} < 0$ \AND $\alpha^{{{k+1}}}_{t_k} < C$}
  		\REPEAT
  		\STATE $\alpha^{{{k+1}}}_{t_k} = \alpha^{{{k+1}}}_{t_k} + \delta$;
  		\STATE Update the balance equation (\ref{eq:constr});
  		\UNTIL { $\alpha^{{{k+1}}}_{t_k} = C$ \OR $g^{{{k+1}}}_{t_k} = 0$ \OR $\exists~x_{s_i} \in \mathcal{S},~\alpha_{s_i} = 0$ \OR $\exists~x_{s_i} \in \mathcal{S},~\alpha_{s_i} = C$ \OR $\exists~x_{b_i} \in \mathcal{B},~g^{b_i} = 0$ \OR $\exists~x_{o_i} \in \mathcal{O},~g^{o_i} = 0$}
  \STATE Support vector pruning for $\mathcal{S}$, such that $\overline{\Xi}_{\mathcal{SS}}^{-1}$ exists;
  \STATE Adjust $\mathcal{B}$ and $\mathcal{O}$;
  \STATE Recalculate $\xi^{\mathcal{S}}_{{t_{k+1}}}$, $\xi^{\mathcal{B}}_{{t_{k+1}}}$, $\xi^{\mathcal{O}}_{{t_{k+1}}}$ and $\Xi_{\mathcal{SS}}$, $\Xi_{\mathcal{BS}}$, $\Xi_{\mathcal{OS}}$ using (\ref{eq:vm});
  \STATE Calculate $\Delta \alpha^{{{k+1}}}_{t_k} = \alpha^{{{k+1}}}_{{t_{k+1}}} -\alpha^{{{k+1}}}_{t_k}$;
  \STATE Calculate $\Delta \alpha^{\mathcal{S}}_{{t_k}}$, $\Delta g^{\mathcal{B}}_{{t_k}}$, $\Delta g^{\mathcal{O}}_{{t_k}}$, $\Delta \alpha^{{{k+1}}}_{t_k}$ and $\Delta g^{{{k+1}}}_{t_k}$ using (\ref{eq:mcp}) and (\ref{eq:comp});
  \STATE Update $\alpha^{\mathcal{S}}_{{t_{k+1}}}$, $g^{\mathcal{B}}_{{t_{k+1}}}$, $g^{\mathcal{O}}_{{t_{k+1}}}$, $\alpha^{k+1}_{{t_{k+1}}}$ and $g^{k+1}_{{t_{k+1}}}$;
  \ENDWHILE
  \STATE Update $w_{{t_{k+1}}}$ based on (\ref{eq:ldc});
\end{algorithmic} 
\end{algorithm}

\begin{remark}
	In Algorithm \ref{ag:rme}, $\delta > 0$ is a adequately small value. The prototype of this algorithm can be found in~\cite{cauwenberghs2001incremental,laskov2006incremental}, and we adopted it for the switching manifold estimation problem in this paper. Note that 7 in Algorithm \ref{ag:rme} is the support vector pruning procedure to guarantee $\overline{\Xi}_{\mathcal{SS}}$ to be non-singular, which will be discussed in Section \ref{sec:svpp}. We will also show that Algorithm \ref{ag:rme} is stable with respect to the convergence of $\alpha^{\mathcal{S}}_{t_k}$, $g^{\mathcal{B}}_{{t_{k}}}$, $g^{\mathcal{O}}_{{t_{k}}}$ as $k$ increases.
\end{remark}

\subsubsection{Support Vector Pruning Procedure}\label{sec:svpp}

It should be noticed that, the $\overline{\Xi}_{\mathcal{SS}}$ is not necessarily non-singular, and if not, (\ref{eq:mcp}) can not be directly used to calculate $\Delta \alpha^{\mathcal{S}}_{t_k}$. Thus conditions for the existence of $\overline{\Xi}_{\mathcal{SS}}^{-1}$ should be investigated.

\begin{lemma}\label{lm:lemma1}
If $\kappa:\mathbb{R}^{n} \times \mathbb{R}^n \rightarrow \mathbb{R}$ is a linear kernel defined as in (\ref{eq:kernel}) and $\mathcal{S} \neq \varnothing$, then $\Xi_{\mathcal{SS}}$ is positive definite if and only if $n_s \leqslant n$ and the support vectors in $\mathcal{S}$ are independent from each other.
\end{lemma}

\begin{proof}
If $\kappa$ is defined as in (\ref{eq:kernel}), then $\Xi_{\mathcal{SS}}$ can be represented as $\Xi_{\mathcal{SS}} = \mathcal{X}_{\mathcal{S}}^T Q \mathcal{X}_{\mathcal{S}}$, where
\begin{equation}\label{eq:xs}
\mathcal{X}_{\mathcal{S}} = \left[~l_{s_1}x_{s_1}~~l_{s_2}x_{s_2}~~\cdots~~l_{s_{n_s}}x_{s_{n_s}} \right].
\end{equation}
It is not difficult to inferred that if $n_s \leqslant n$ and $x_{s_1}$, $x_{s_2}$, $\cdots$, $x_{s_{n_s}}$ are independent from each other, then $\mathcal{X}_{\mathcal{S}}$ is non-singular. Thus $\Xi_{\mathcal{SS}}$ is non-singular and positive definite. Note that the inverse proposition also holds, which proves the lemma.
\end{proof}

\begin{lemma}\label{lm:lemma2}
If $\Xi_{\mathcal{SS}}^{-1}$ exists and $\Xi_l = l_{\mathcal{S}}^T \Xi_{\mathcal{SS}}^{-1} l_{\mathcal{S}}$ is non-singular, then the augmented matrix $\overline{\Xi}_{\mathcal{SS}}$ as in (\ref{eq:comp}) is non-singular and its inverse reads
\begin{equation}
\overline{\Xi}_{\mathcal{SS}}^{-1} = \left[ \begin{array}{cc}
-\Xi_l^{-1} & \Xi_l^{-1} l_{\mathcal{S}}^T \Xi_{\mathcal{SS}}^{-1} \\
\Xi_{\mathcal{SS}}^{-1} l_{\mathcal{S}} \Xi_l^{-1} & \Xi_{\mathcal{SS}}^{-1} + \Xi_{\mathcal{SS}}^{-1} l_{\mathcal{S}} \Xi_l^{-1} l^T_{\mathcal{S}} \Xi^{-1}_{\mathcal{SS}}
\end{array} \right].
\end{equation}
\end{lemma}

\begin{proof}
By calculating $\overline{\Xi}^{-1}_{\mathcal{SS}} \overline{\Xi}_{\mathcal{SS}}$ and $\overline{\Xi}_{\mathcal{SS}} \overline{\Xi}^{-1}_{\mathcal{SS}}$ we have
\begin{equation}
\overline{\Xi}^{-1}_{\mathcal{SS}} \overline{\Xi}_{\mathcal{SS}} = \overline{\Xi}_{\mathcal{SS}} \overline{\Xi}^{-1}_{\mathcal{SS}} =I,
\end{equation}
which proves the lemma.
\end{proof}

\begin{lemma}\label{lm:lemma3}
If $\kappa$ is defined as in (\ref{eq:kernel}), then the SVM model fitting the training set $\mathcal{T}_{t_k}$ is uniquely determined only if $\mathrm{rank} \left( \mathcal{X}_{\mathcal{S}} \right) =n$, with $\mathcal{X}_{\mathcal{S}}$ defined in (\ref{eq:xs}).
\end{lemma}
\begin{proof}
According to (\ref{eq:ldc}) and (\ref{eq:gvl}) with $b=0$, we have $l_{s_i}f(x_{s_i})  = 1$, $\forall~x_{s_i} \in \mathcal{S}$. Then the trained SVM model is parameterized by $w$ which is the solution of
\begin{equation}
	w^T x_{s_i}l_{s_i} = 1,~\forall~x_{s_i} \in \mathcal{S},
\end{equation}
which leads to
\begin{equation}\label{eq:weq}
	w^T \mathcal{X}_{\mathcal{S}} = 1_n,
\end{equation}
where $1_n \in \mathbb{R}^n$ is the vector with full $1$ elements. Thus it is known that the solution $w$ in (\ref{eq:weq}) is unique only if $\mathrm{rank} \left( \mathcal{X}_{\mathcal{S}} \right) = n$, which proves the lemma.
\end{proof}

Lemma \ref{lm:lemma1} and Lemma \ref{lm:lemma2} give the sufficient conditions for $\overline{\Xi}_{\mathcal{SS}}$ to be non-singular, and Lemma \ref{lm:lemma3} reveals the necessary condition that the SVM model is uniquely determined.
As a result, the non-singularity of $\overline{\Xi}_{\mathcal{SS}}$ and uniqueness of SVM model are both guaranteed, if $\mathrm{rank} \left( \overline{\Xi}_{\mathcal{SS}} \right) = n_s = n$ holds and $l_{\mathcal{S}}^T \Xi_{\mathcal{SS}}^{-1} l_{\mathcal{S}}$ is non-singular. Based on this, we propose the following \textit{support vector pruning scheme} as in 7 of Algorithm \ref{ag:rme},

(1). If $\mathrm{rank} \left( \overline{\Xi}_{\mathcal{SS}} \right) = n_s = n$, then $\overline{\Xi}_{\mathcal{SS}}^{-1}$ exists, and (\ref{eq:mcp}) is valid if $l_{\mathcal{S}}^T \Xi_{\mathcal{SS}}^{-1} l_{\mathcal{S}}$ is non-singular;

(2). If $\mathrm{rank} \left( \overline{\Xi}_{\mathcal{SS}} \right) < n_s$, find out redundant support vectors in $\mathcal{S}$ and remove them, such that $n_s$ linearly independent support vectors remain and check (3);

(3). If $\mathrm{rank} \left( \overline{\Xi}_{\mathcal{SS}} \right) = n_s < n$, select $n-n_s$ \textit{pseudo} support vectors on the current SVM hyper-plane pair $g(x) = 0$ as in (\ref{eq:gvl}), such that $\mathrm{rank} \left( \overline{\Xi}_{\mathcal{SS}} \right) = n_s = n$ holds and (\ref{eq:mcp}) is valid if $l_{\mathcal{S}}^T \Xi_{\mathcal{SS}}^{-1} l_{\mathcal{S}}$ is non-singular. Note that these \textit{pseudo} support vectors should be eliminated when $\mathcal{S}$ varies.

If $l_{\mathcal{S}}^T \Xi_{\mathcal{SS}}^{-1} l_{\mathcal{S}}$ is singular, then (\ref{eq:mcp}) is equivalent to the left equation in (\ref{eq:mat1}), and the constraint as the right equation in (\ref{eq:mat1}) is redundant. Thus in this case, $\Delta \alpha^{S}_{t_k}$ can be calculated by (\ref{eq:mat1}).

\subsection{Stability Analysis of Manifold Estimation}
In this chapter, we discuss the stability of Algorithm \ref{ag:rme}. First of all, we investigate the sufficient conditions of the invariance of $\mathcal{S}$, $\mathcal{B}$ and $\mathcal{O}$ during the incremental tuning. If the support vector pruning process in Section \ref{sec:svpp} is proceeded, then $\mathrm{rank} \left( \mathcal{X}_{\mathcal{S}} \right) = n_{s} = n$ holds, and it is known then that any samples in $\Omega$ can be represented as the linear combination of the support vectors, and so does $x_{{t_{k+1}}}$, i.e.
\begin{equation}
x_{{t_{k+1}}} = \sum \limits^{n_s}_{i=1} \sigma^{s_i} x_{s_i},
\end{equation}
which leads to
\begin{equation}\label{eq:sigs}
l_{k+1}x_{{t_{k+1}}} = \sum \limits^{n_s}_{i=1} l_{k+1} \sigma^{s_i} x_{s_i} = \sum \limits^{n_s}_{i=1} \left( l_{k+1} \sigma^{s_i} l_{s_i} \right) l_{s_i} x_{s_i} = \mathcal{X}_{\mathcal{S}} \sigma^{\mathcal{S}}_{{t_{k+1}}},
\end{equation}
where $\sigma^{\mathcal{S}}_{{t_{k+1}}} = \left[l_{k+1} l_{s_1} \sigma^{s_1}~l_{k+1} l_{s_2}\sigma^{s_2}~\cdots~l_{k+1} l_{s_{n_s}}\sigma^{s_{n_s}} \right]^T \in \mathbb{R}^{n_s}$ is the coefficient vector of $x_{{t_{k+1}}}$.

\begin{lemma}\label{lm:lemma4}
If $\sigma^{\mathcal{S}}_{{t_{k+1}}}$ is defined as in (\ref{eq:sigs}) and the following condition holds,
\begin{equation}\label{eq:suff}
\max \limits_i \left( \alpha^{\mathcal{S}}_{{t_k}} - C \sigma^{\mathcal{S}}_{{t_{k+1}}} \right)_i < C,~\min \limits_i \left( \alpha^{\mathcal{S}}_{{t_k}} - C \sigma^{\mathcal{S}}_{{t_{k+1}}} \right)_i >0,
\end{equation}
where $(\cdot)_i$ denotes the $i$-th element of the vector, then sets $\mathcal{S}$, $\mathcal{B}$ and $\mathcal{O}$ are invariant during the incremental tuning for $x_{{t_{k+1}}}$.
\end{lemma}

\begin{proof}
Combining (\ref{eq:mcp}) and (\ref{eq:vm}), and substituting (\ref{eq:sigs}), $\xi^{\mathcal{S}}_{{t_{k+1}}}$ can be represented as
\begin{equation}\label{eq:xis}
\xi^{\mathcal{S}}_{{t_{k+1}}} = \mathcal{X}_{\mathcal{S}}^T Q x_{{t_{k+1}}}l_{k+1}  = \mathcal{X}_{\mathcal{S}}^T Q \mathcal{X}_{\mathcal{S}} \sigma^{\mathcal{S}}_{{t_{k+1}}} = \Xi_{\mathcal{SS}} \sigma^{\mathcal{S}}_{{t_{k+1}}}.
\end{equation}
Substituting (\ref{eq:xis}) to (\ref{eq:mat1}), we have
\begin{equation}
\Delta \alpha^{\mathcal{S}}_{{t_k}} = - \Xi_{\mathcal{SS}}^{-1} \Xi_{\mathcal{SS}} \sigma^{\mathcal{S}}_{{t_{k+1}}} \Delta \alpha^{{{k+1}}}_{t_k} = - \sigma^{\mathcal{S}}_{{t_{k+1}}} \Delta \alpha^{{{k+1}}}_{t_k},
\end{equation}
and the updated weight vector $\alpha^{\mathcal{S}}_{{t_{k+1}}}$ reads
\begin{equation}\label{eq:alup}
\alpha^{\mathcal{S}}_{{t_{k+1}}} = \alpha^{\mathcal{S}}_{{t_k}} - \sigma^{\mathcal{S}}_{{t_{k+1}}} \Delta \alpha^{{{k+1}}}_{t_k}.
\end{equation}
If (\ref{eq:suff}) holds, we have
\begin{equation}
0< \left( \alpha^{\mathcal{S}}_{t_k} \right)_i - C \left( \sigma^{\mathcal{S}}_{t_{k+1}} \right)_i < C,~\forall~1 \leqslant i \leqslant n_s.
\end{equation}
Considering $0 \leqslant \Delta \alpha^{{{k+1}}}_{t_k} \leqslant C$, then for each $i$, $s_1 \leqslant i \leqslant s_n$, if $\left( \sigma^{\mathcal{S}}_{t_{k+1}} \right)_i > 0$
\begin{equation}
0 < \left( \alpha^{\mathcal{S}}_{t_{k}} \right)_i - C \left( \sigma^{\mathcal{S}}_{t_{k+1}} \right)_i \leqslant \left( \alpha^{\mathcal{S}}_{{t_k}} \right)_i - \left( \sigma^{\mathcal{S}}_{{t_{k+1}}} \Delta \alpha^{{{k+1}}}_{t_k} \right)_i \leqslant \left( \alpha^{\mathcal{S}}_{{t_k}} \right)_i < C,
\end{equation}
while if $\left( \sigma^{\mathcal{S}}_{t_{k+1}} \right)_i < 0$,
\begin{equation}
0 < \left( \alpha^{\mathcal{S}}_{{t_k}} \right)_i \leqslant \left( \alpha^{\mathcal{S}}_{{t_k}} \right)_i - \left( \sigma^{\mathcal{S}}_{{t_{k+1}}} \Delta \alpha^{{{k+1}}}_{t_k} \right)_i \leqslant \left( \alpha^{\mathcal{S}}_{t_{k}} \right) - C \left( \sigma^{\mathcal{S}}_{t_{k+1}} \right)_i < C.
\end{equation}
Therefore
\begin{equation}
0 < \left( \alpha^{\mathcal{S}}_{{t_k}} \right)_i - \left( \sigma^{\mathcal{S}}_{{t_{k+1}}} \Delta \alpha^{t_{k+1}}_{t_k} \right)_i = \left( \alpha^{\mathcal{S}}_{{t_{k+1}}} \right)_i < C
\end{equation}
is obtained for all $1 \leqslant i \leqslant n_s$, which means that none of the support vectors in $\mathcal{S}$ gets saturated. Thus the invariance of $\mathcal{S}$, $\mathcal{B}$ and $\mathcal{O}$ is proven.
\end{proof}

\begin{corollary}\label{cr:co1}
If $\sigma^{\mathcal{S}}_{{t_{k+1}}}$ is defined as in (\ref{eq:sigs}) and the following condition holds for all $r \geqslant {k+1}$,
\begin{equation}\label{eq:coro}
\max \limits_i \left( \alpha^{\mathcal{S}}_{{t_k}} - C \sum \limits_{j=k}^{r-1} \sigma^{\mathcal{S}}_{t_{j+1}} \right)_i < C,~\min \limits_i \left( \alpha^{\mathcal{S}}_{{t_k}} - C \sum \limits_{j=k}^{r-1}  \sigma^{\mathcal{S}}_{t_{j+1}} \right)_i >0,
\end{equation}
then the sample sets $\mathcal{S}$, $\mathcal{B}$ and $\mathcal{O}$ are invariant for all incremental tuning $\forall r \geqslant k+1$.
\end{corollary}

\begin{proof}
Since we have
\begin{equation}
\alpha^{\mathcal{S}}_{t_r} = \alpha^{\mathcal{S}}_{t_k} - \sum \limits_{j={{k}}}^{r-1} \sigma^{\mathcal{S}}_{t_{j+1}} \Delta \alpha^{{j+1}}_{t_j},~r \geqslant {k+1},
\end{equation}
and $0 \leqslant \Delta \alpha^{{{j+1}}}_{t_j} \leqslant C$, $\forall~{k} \leqslant j \leqslant r-1$, then there exist $\bar{\alpha} \in \mathbb{R}^{n_s}$, 
\begin{equation}
0 < \min \limits_{j} \left( \Delta \alpha^{{j+1}}_{t_j} \right)_i \leqslant \left( \bar{\alpha} \right)_i \leqslant \max \limits_{j} \left( \Delta \alpha^{{j+1}}_{t_j} \right)_i < C,~\forall~ 1 \leqslant i \leqslant n_s,
\end{equation}
such that
\begin{equation}
\alpha^{\mathcal{S}}_{t_r} = \alpha^{\mathcal{S}}_{t_k} - \sum \limits_{j={{k}}}^{r-1} \sigma^{\mathcal{S}}_{t_{j+1}} \Delta \alpha^{{j+1}}_{t_j} = \alpha^{\mathcal{S}}_{t_k} -  \left( \sum \limits_{j={{k}}}^{r-1} \sigma^{\mathcal{S}}_{t_{j+1}} \right) \bar{\alpha}.
\end{equation}
With similar arguments as Lemma \ref{lm:lemma4}, it is not difficult to prove that if (\ref{eq:coro}) holds, then $\mathcal{S}$, $\mathcal{B}$ and $\mathcal{O}$ are invariant.
\end{proof}

\begin{theorem}\label{cr:co2}
If the sample sequence $\mathcal{X}_{\tau}$ satisfies that
\begin{equation}\label{eq:cka}
\kappa(x_{{t_{k+1}}},x_{{t_{k+1}}}) \leqslant \epsilon \kappa(x_{{t_k}},x_{{t_k}}),~0 < \epsilon < 1,~\forall~{t_k} \geqslant 0,
\end{equation}
and the following condition holds $\forall~1 \leqslant i \leqslant n_s$,
\begin{equation}\label{eq:coro2}
\max \limits_i \left( \alpha^{\mathcal{S}}_{{t_k}} \right)_i < C \left( 1- \frac{ \left(\sqrt{\epsilon}\right)^{{t_{k+1}}}}{1- \sqrt{\epsilon}} \sqrt{\frac{\kappa_{t_0}}{\lambda_{\mathrm{m}}^{\Xi}}} \right),~\min \limits_i \left( \alpha^{\mathcal{S}}_{{t_k}} \right)_i >C\frac{ \left(\sqrt{\epsilon}\right)^{{t_{k+1}}}}{1-\sqrt{\epsilon}} \sqrt{\frac{\kappa_{t_0}}{\lambda_{\mathrm{m}}^{\Xi}}},
\end{equation}
where $\kappa_{t_0} = \kappa \left(x_{t_0},x_{t_0} \right)$ is the kernel function of the initial sample $x_{t_0}$, $\lambda_{\mathrm{m}}^{\Xi}$ is the minimum eigenvalue of $\Xi_{\mathcal{SS}}$, and $\varrho_{{t_{k+1}}} =\left\| \sigma^{\mathcal{S}}_{{t_{k+1}}} \right\|_{\infty} \left/ \left\|\sigma^{\mathcal{S}}_{{t_{k+1}}} \right\|_2 \right.$ denotes the ratio of the norms of $\sigma^{\mathcal{S}}_{{t_{k+1}}}$, then $\mathcal{S}$, $\mathcal{B}$ and $\mathcal{O}$ are invariant for all incremental tuning since after sample $x_{t_{k+1}}$.
\end{theorem}

\begin{proof}
For each element $\left( \sigma^{\mathcal{S}}_{{t_{k+1}}} \right)_i$ in $\sigma^{\mathcal{S}}_{{t_{k+1}}}$, $1 \leqslant i \leqslant {n_s}$, we have
\begin{equation}\label{eq:cor21}
\begin{split}
-\max \limits_i  \left| \left( \sigma^{\mathcal{S}}_{{t_{k+1}}} \right)_i \right| &\leqslant -  \left| \left( \sigma^{\mathcal{S}}_{{t_{k+1}}} \right)_i \right| \leqslant \left( \sigma^{\mathcal{S}}_{{t_{k+1}}} \right)_i \\ &\leqslant   \left| \left( \sigma^{\mathcal{S}}_{{t_{k+1}}} \right)_i \right| \leqslant  \max \limits_i  \left| \left( \sigma^{\mathcal{S}}_{{t_{k+1}}} \right)_i \right|,
\end{split}
\end{equation}
where $|\cdot|$ denotes the absolute value of a scalar. Also,
\begin{equation}\label{eq:cor24}
\max \limits_i  \left| \left( \sigma^{\mathcal{S}}_{{t_{k+1}}} \right)_i \right| = \left\| \sigma^{\mathcal{S}}_{{t_{k+1}}} \right\|_{\infty} = \varrho_{{t_{k+1}}} \left\|\sigma^{\mathcal{S}}_{{t_{k+1}}} \right\|_2,
\end{equation}
and note that
\begin{equation}
\frac{1}{\sqrt{n_s}} \leqslant \varrho_{t_{k+1}} \leqslant 1.
\end{equation}
Rewriting (\ref{eq:kernel}) we get
\begin{equation}\label{eq:cor22}
\kappa \left(x_{{t_{k+1}}}, x_{{t_{k+1}}} \right) =\sigma^{\mathcal{S}T}_{{t_{k+1}}} \mathcal{X}_{\mathcal{S}}^T Q \mathcal{X}_{\mathcal{S}} \sigma^{\mathcal{S}}_{{t_{k+1}}} = \sigma^{\mathcal{S}T}_{{t_{k+1}}} \Xi_{\mathcal{SS}} \sigma^{\mathcal{S}}_{{t_{k+1}}} \geqslant \lambda_{\mathrm{m}}^{\Xi} \left\|\sigma^{\mathcal{S}}_{{t_{k+1}}} \right\|^2_2,
\end{equation}
and additionally (\ref{eq:cka}) leads to
\begin{equation}\label{eq:cor23}
\kappa \left(x_{{t_{k+1}}},x_{{t_{k+1}}} \right) \leqslant \epsilon^{{t_{k+1}}} \kappa  \left(x_{0},x_{0} \right) =  \epsilon^{{t_{k+1}}}  \kappa_{t_0}.
\end{equation}
Substituting (\ref{eq:cor22}) and (\ref{eq:cor23}) to (\ref{eq:cor24}) we have
\begin{equation}\label{eq:cor25}
\max \limits_i  \left| \left( \sigma^{\mathcal{S}}_{{t_{k+1}}} \right)_i \right| \leqslant \left( \sqrt{\epsilon} \right)^{{t_{k+1}}} \varrho_{{t_{k+1}}} \sqrt{\frac{\kappa_{t_0}}{\lambda^{\Xi}_{\mathrm{m}}}} = \left( \sqrt{\epsilon} \right)^{{t_{k+1}}}  \sqrt{\frac{\kappa_{t_0}}{\lambda^{\Xi}_{\mathrm{m}}}},
\end{equation}
and the summary of (\ref{eq:cor25}) from ${{k+1}}$ to $r$, $\forall r \geqslant k+1$ reads
\begin{equation}\label{eq:cor26}
\begin{split}
\sum \limits_{j=k}^{r-1} \max \limits_i  \left| \left( \sigma^{\mathcal{S}}_{{t_{j+1}}} \right)_i \right| &= \sum \limits_{j=k}^{+ \infty} \max \limits_i  \left| \left( \sigma^{\mathcal{S}}_{{t_{j+1}}} \right)_i \right| \\ & \leqslant  \sum \limits_{j={{k}}}^{+\infty}  \left( \sqrt{\epsilon} \right)^{{t_{k+1}}} \sqrt{\frac{\kappa_{t_0}}{\lambda^{\Xi}_{\mathrm{m}}}}  = \frac{\left( \sqrt{\epsilon} \right)^{{t_{k+1}}}}{1 -\sqrt{\epsilon}}  \sqrt{\frac{\kappa_{t_0}}{\lambda^{\Xi}_{\mathrm{m}}}}.
\end{split}
\end{equation}
Therefore, combining (\ref{eq:cor21}) and (\ref{eq:cor26}) we obtain
\begin{equation}
\begin{split}
-\frac{\left( \sqrt{\epsilon} \right)^{{t_{k+1}}}}{1 -\sqrt{\epsilon}}  \sqrt{\frac{\kappa_{t_0}}{\lambda^{\Xi}_{\mathrm{m}}}}& \leqslant - \sum \limits_{j={{k}}}^{r-1} \max \limits_i  \left| \left( \sigma^{\mathcal{S}}_{{t_{j+1}}} \right)_i \right| \leqslant \sum \limits_{j={{k}}}^{r-1}  \left( \sigma^{\mathcal{S}}_{{t_{j+1}}} \right)_i \\
&\leqslant \sum \limits_{j={{k}}}^{r-1} \max \limits_i  \left| \left( \sigma^{\mathcal{S}}_{{t_{j+1}}} \right)_i \right|  \leqslant \frac{\left( \sqrt{\epsilon} \right)^{{t_{k+1}}}}{1 -\sqrt{\epsilon}}  \sqrt{\frac{\kappa_{t_0}}{\lambda^{\Xi}_{\mathrm{m}}}}.
\end{split}
\end{equation}
According to (\ref{eq:coro}), we have 
\begin{equation}\label{eq:cof1}
\max \limits_i \left( \alpha^{\mathcal{S}}_{{t_k}} \right)_i + \frac{C \left( \sqrt{\epsilon} \right)^{{t_{k+1}}}}{1 -\sqrt{\epsilon}}  \sqrt{\frac{\kappa_0}{\lambda^{\Xi}_{\mathrm{m}}}} < C,~\min \limits_i \left( \alpha^{\mathcal{S}}_{{t_k}} \right)_i - \frac{C \left( \sqrt{\epsilon} \right)^{{t_{k+1}}}}{1 -\sqrt{\epsilon}}  \sqrt{\frac{\kappa_0}{\lambda^{\Xi}_{\mathrm{m}}}} >0.
\end{equation}
Similar to Corollary \ref{cr:co1}, we have
\begin{equation}\label{eq:cof2}
\alpha^{\mathcal{S}}_{t_r} = \alpha^{\mathcal{S}}_{t_k} - \sum \limits_{j={{k}}}^{r-1} \sigma^{\mathcal{S}}_{t_{j+1}} \Delta \alpha^{{j+1}}_{t_j} = \alpha^{\mathcal{S}}_{t_k} -  \left( \sum \limits_{j={{k}}}^{r-1} \sigma^{\mathcal{S}}_{t_{j+1}} \right) \bar{\alpha}.
\end{equation}
Combining (\ref{eq:cof1}) and (\ref{eq:cof2}), we have
\begin{equation}
\alpha^{\mathcal{S}}_{t_r} \leqslant \max \limits_i \left( \alpha^{\mathcal{S}}_{t_k} \right)_i + C  \sum \limits_{j={{k}}}^{r-1} \max \limits_i  \left| \left( \sigma^{\mathcal{S}}_{{t_{j+1}}} \right)_i \right| \leqslant C,
\end{equation}
and on the other hand
\begin{equation}
\alpha^{\mathcal{S}}_{t_r} \geqslant \min \limits_i \left( \alpha^{\mathcal{S}}_{t_k} \right)_i - C  \sum \limits_{j={{k}}}^{r-1} \max \limits_i  \left| \left( \sigma^{\mathcal{S}}_{{t_{j+1}}} \right)_i \right| \geqslant 0,
\end{equation}
for all $r \geqslant k+1$, which proves the corollary.
\end{proof}

\begin{remark}
Lemma \ref{lm:lemma4} and Corollary \ref{cr:co1} and \ref{cr:co2} reveal the conditions that guarantee the invariance of $\mathcal{S}$, $\mathcal{B}$ and $\mathcal{O}$ during the sequential sampling of the system. Note that the condition (\ref{eq:cka}) is guaranteed by the exponential stability condition in Assumption \ref{as:uasum}, if the kernel function $\kappa \left(x_{t_k}, x_{t_k}\right)$ has the same form as the Lyapunov function $V_{t_k}$, i.e. $Q$ is set as $Q=P$, where $P$ is defined in Assumption \ref{as:uasum}. Since $\sqrt{\epsilon}$, $\kappa_{t_0}$ and $\lambda^{\Xi}_{\mathrm{m}}$ respectively depend on predefined constant, initial sample and support vectors, it can be inferred that the condition (\ref{eq:coro2}) holds when ${t_k}$ gets adequately large and $\left( \sqrt{\epsilon} \right)^{t_{k+1}}$ is adequately small.
\end{remark}

\begin{proposition}
If sample sets $\mathcal{S}$, $\mathcal{B}$ and $\mathcal{O}$ are invariant for new samples $x_{t_r}$, $\forall~r \geqslant k+1$, then the SVM model parameterized by $w_{t_r}$ is also invariant.
\end{proposition}

This proposition describes a important feature of SVM that the classifier is only determined by support vectors. Also considering Theorem \ref{cr:co2}, it is noticed that the incremental SVM model training can be conducted by finite amount of samples. Further, we present the following stability proof of Algorithm \ref{ag:rme}.

\begin{theorem}\label{th:theo}
    If sample sets $\mathcal{S}$, $\mathcal{B}$ and $\mathcal{O}$ are invariant during the incremental tuning as ${k} \rightarrow + \infty$, then Algorithm \ref{ag:rme} guarantees that the support vector weights $\alpha^{\mathcal{S}}_{t_k}$ converge to the ideal SVM weight $\alpha^{\mathcal{S}}_{*}$ that fits the infinitely sampled state sequence $\mathcal{X}_{\infty}$ if the following conditions hold,
    \begin{equation}\label{eq:thcond2}
        \frac{3}{2} \kappa \left(x_{k+1},x_{{t_{k+1}}} \right) < \min \limits_j \kappa \left(x_{{t_{k+1}}},x_{s_j} \right)
    \end{equation}
    and
    \begin{equation}\label{eq:thcond1}
        l_{k+1}l_i \left( \kappa(x_{{t_{k+1}}},x_{i}) - \overline{\kappa}^{{t_{k+1}}}_i \right) \leqslant 0,~\forall~i \geqslant {t_{k+1}},
    \end{equation}
    where $x_{t_i}$, $i > {{k+1}}$ are the future samples after ${{k+1}}$, and $\overline{\kappa}^{{{k+1}}}_i$ is a real scalar
    \begin{equation}
        \min \limits_{j} \kappa \left(x_{{t_{k+1}}},x_{s_j} \right) \leqslant \overline{\kappa}^{{{k+1}}}_i \leqslant  \max \limits_{j} \kappa \left(x_{{t_{k+1}}},x_{s_j} \right)
    \end{equation}
     such that
	\begin{equation}\label{eq:kpavr}
		\overline{\kappa}^{{{k+1}}}_i = \left( \sum \limits_{j=1}^{n_s} \kappa \left( x_{{t_{k+1}}}, x_{s_j} \right) l_{s_j} \Delta \alpha^{s_j}_{i} \right) \left/ \left(\sum \limits_{j=1}^{n_s}  l_{s_j} \Delta \alpha^{s_j}_{i} \right. \right).
	\end{equation}	    
\end{theorem}

\begin{proof}
	Since $\Xi_{\mathcal{SS}}$ is positive definite according to Lemma \ref{lm:lemma1}, we define the following positive definite function
    \begin{equation}
        V^{\mathcal{S}}_{t_k} = \frac{1}{2} \left( \alpha^{\mathcal{S}}_{t_k} - \alpha^{\mathcal{S}}_* \right) ^T \Xi_{\mathcal{SS}} \left( \alpha^{\mathcal{S}}_{t_k} - \alpha^{\mathcal{S}}_* \right),
    \end{equation}
	where $\alpha^{\mathcal{S}}_* = \left[ \alpha^{s_1}_*,~\alpha^{s_2}_*,~\cdots,~\alpha^{s_n}_* \right]^T$ represents the ideal weights of the SVM support vectors corresponding to $w_*$. Thus the increment $\Delta V^{\mathcal{S}}_{t_k} = V^{\mathcal{S}}_{{t_{k+1}}} - V^{\mathcal{S}}_{t_k}$ reads
    \begin{equation}\label{eq:pf1}
        \Delta V^{\mathcal{S}}_{t_k} = \left( \alpha^{\mathcal{S}}_{t_k} - \alpha^{\mathcal{S}}_* \right)^T  \hspace{-0.1cm} \Xi_{\mathcal{SS}} \Delta \alpha^{\mathcal{S}}_{t_k} + \frac{1}{2} \left( \Delta \alpha^{\mathcal{S}}_{t_k} \right)^T \hspace{-0.1cm} \Xi_{\mathcal{SS}} \Delta \alpha^{\mathcal{S}}_{t_k}.
    \end{equation}
	Substituting (\ref{eq:mat1}) to (\ref{eq:pf1}), we have 
	\begin{equation}\label{eq:lya1}
	\frac{1}{2} \left( \Delta \alpha^{\mathcal{S}}_{t_k}  \right)^T \Xi_{\mathcal{SS}} \Delta \alpha^{\mathcal{S}}_{t_k}  = \frac{1}{2} \left( \Delta \alpha_{{t_k}}^{k+1} \right)^2 \sigma_{\mathcal{S}}^T \Xi_{\mathcal{SS}} \sigma_{\mathcal{S}} = \frac{1}{2} \left( \Delta \alpha_{{t_k}}^{k+1} \right)^2 \kappa \left( x_{{t_{k+1}}}, x_{{t_{k+1}}} \right),
	\end{equation}
	and
	\begin{equation}\label{eq:fnp1}
		\left( \alpha^{\mathcal{S}}_{t_k} - \alpha^{\mathcal{S}}_* \right)^T \Xi_{\mathcal{SS}} \Delta \alpha^{\mathcal{S}}_{t_k} = -\Delta \alpha_{{t_k}}^{k+1} \hspace{-0.05cm} \left( \alpha^{\mathcal{S}}_{t_k} - \alpha^{\mathcal{S}}_* \right)^T \xi^{\mathcal{S}}_{{t_{k+1}}}. 
	\end{equation}	    
    From the definition of $g(x)$ as in (\ref{eq:gvl}) we have
    \begin{equation}\label{eq:gkplus}
    g_{t_k}^{k+1} = l_{k+1} \left( \sum \limits_{i=1}^{{k}} \alpha^i_{t_k} l_i \kappa(x_{t_i},x_{{t_{k+1}}}) \right) -1,~\forall~x_{t_i} \in \mathcal{T}.
    \end{equation}
    Since $\alpha^i = C$, $\forall~x_i \in \mathcal{B}$ and $\alpha^i = 0$, $\forall~x_i \in \mathcal{O}$, (\ref{eq:gkplus}) is further expended as
    \begin{equation}\label{eq:gkplus1}
    g_{t_k}^{k+1} = l_{k+1} \left( \sum \limits_{i = 1}^{n_s}   \alpha^{s_i}_{t_k} l_{s_i}  \kappa(x_{s_i},x_{{t_{k+1}}})  + C \sum \limits_{i=1}^{n_b} l_{b_i} \kappa(x_{b_i},x_{{t_{k+1}}}) \right) - 1. 
    \end{equation}
    Similarly we have
    \begin{equation}\label{eq:gkplusstar}
    g_*^{k+1} = l_{k+1} \left( \sum \limits_{i = 1}^{n_s} \right.  \alpha_*^{s_i} l_{s_i}  \kappa(x_{s_i},x_{{t_{k+1}}}) + \left. C \sum \limits_{i=1}^{n_b} l_{b_i} \kappa(x_{b_i},x_{{t_{k+1}}}) \right) - 1, 
    \end{equation}  
	where $g^{k+1}_*$ denotes the ideal $g$ value of $x_{{t_{k+1}}}$ corresponding to the ideal SVM model. By calculating (\ref{eq:gkplus1})-(\ref{eq:gkplusstar}) we have    
	\begin{equation}\label{eq:fnp2}
	g_{t_k}^{k+1} - g_*^{k+1} = l_{k+1} \sum \limits_{i = 1}^{n_s} \left( \alpha^{s_i}_{t_k} - \alpha_*^{s_i} \right) l_{s_i}  \kappa(x_{s_i},x_{{t_{k+1}}})  = \left( \alpha^{\mathcal{S}}_{t_k} - \alpha^{\mathcal{S}}_* \right)^T \xi^{\mathcal{S}}_{{t_{k+1}}}.
	\end{equation}	    
	Additionally,
	\begin{equation}\label{eq:gg}
		g_{t_k}^{k+1} - g_*^{k+1} = \sum \limits_{i=k}^{+ \infty} \left( g_{t_i}^{k+1} - g_{t_{i+1}}^{k+1} \right) = -\sum \limits_{i=k}^{+ \infty} \Delta g_{t_i}^{k+1},
	\end{equation}		
	where $g^{k+1}_{t_i}$ and $g^{k+1}_{t_{i+1}}$ are the $g$ values of $x_{{t_{k+1}}}$ before and after the $i$-th incremental tuning, and $\Delta g^{k+1}_{t_k} = g^{k+1}_{t_{i+1}} - g^{k+1}_{t_i}$. According to the incremental KKT condition as in (\ref{eq:constr}) we have
	\begin{equation}\label{eq:gvi}
	\begin{split}
	\Delta g_{t_i}^{k+1} = l_{k+1}l_{i} \kappa \left( x_{{t_{k+1}}},x_{t_i} \right) & \Delta \alpha^{i+1}_{t_i} +  \sum \limits_{j=1}^{n_s} l_{k+1} l_{s_j} \kappa \left( x_{{t_{k+1}}}, x_{s_j} \right) \Delta \alpha^{s_j}_{t_i} \\
     -l_{k+1} l_i \Delta \alpha^{i+1}_{t_i} &= \sum \limits_{j=1}^{n_s} l_{k+1} l_{s_j} \Delta \alpha^{s_j}_{t_i},~ \forall~i \geqslant k+1.
    \end{split}
    \end{equation}
    where $\Delta \alpha^{i+1}_{t_i}$ and $\Delta \alpha^{s_j}_{t_i}$ are respectively the incremental $\alpha$ weights of sample $x_{t_i}$ and support vector $x_{s_j}$ in the $i$-th incremental tuning.
    Substituting (\ref{eq:kpavr}) to (\ref{eq:gvi}) we have
    \begin{equation}\label{eq:sok}
    \Delta g^{k+1}_{t_i} = l_{k+1}l_i \left( \kappa(x_{{t_{k+1}}},x_{t_i}) - \overline{\kappa}^{k+1}_i \right) \Delta \alpha^{i+1}_{t_i} \leqslant 0,
    \end{equation}
    due to $\Delta \alpha^{i+1}_{t_i} \geqslant 0$. Therefore the summary term in (\ref{eq:gg}) reads
    \begin{equation}\label{eq:fnp4}
    \sum \limits_{i={k}}^{+ \infty} \Delta g_{t_i}^{{t_{k+1}}} = \sum \limits_{i=k+1}^{+ \infty} \Delta g_{t_i}^{{t_{k+1}}} + \Delta g_{t_k}^{k+1} \leqslant \Delta g_{t_k}^{k+1}. 
    \end{equation}
    Specially for $i=k$, (\ref{eq:sok}) leads to
    \begin{equation}\label{eq:fnp5}
    \begin{split}
    \Delta g^{k+1}_{t_k} &= l^2_{k+1} \left(\kappa \left( x_{{t_{k+1}}},x_{{t_{k+1}}}\right)  - \overline{\kappa}^{k+1}_{t_k} \right) \Delta \alpha^{{{k+1}}}_{t_k} \\ & \leqslant \left( \kappa(x_{{t_{k+1}}},x_{t_{k+1}}) - \min \limits_j \kappa \left(x_{{t_{k+1}}},x_{s_j} \right) \right) \Delta \alpha_{{t_k}}^{k+1},
    \end{split}
    \end{equation}
    Substituting (\ref{eq:fnp4}) to (\ref{eq:fnp5}) we obtain
    \begin{equation}\label{eq:fnp3}
    \sum \limits_{i=k}^{+ \infty} \Delta g_{t_i}^{k+1}  \leqslant  \left( \kappa(x_{{t_{k+1}}},x_{t_{k+1}}) - \min \limits_j \kappa \left(x_{{t_{k+1}}},x_{s_j} \right) \right) \Delta \alpha_{{t_k}}^{k+1},
    \end{equation}
    and combining (\ref{eq:fnp1}), (\ref{eq:fnp2}), (\ref{eq:gg}) and (\ref{eq:fnp3}), we have
    \begin{equation}\label{eq:lya2}
    \begin{split}
    \left( \alpha^{\mathcal{S}}_{t_k} - \alpha^{\mathcal{S}}_* \right)^T& \Xi_{\mathcal{SS}} \Delta \alpha^{\mathcal{S}}_{t_k} = \Delta \alpha^{{{k+1}}}_{t_k} \left(\sum \limits_{i=k}^{+ \infty} \Delta g_{t_i}^{k+1} \right) \\ & \leqslant \left( \kappa(x_{{t_{k+1}}},x_{t_{k+1}}) - \min \limits_j \kappa \left(x_{{t_{k+1}}},x_{s_j} \right) \right) \left( \Delta \alpha_{{t_k}}^{k+1} \right)^2.
    \end{split}
    \end{equation}
    Substituting (\ref{eq:lya1}) and (\ref{eq:lya2}) to (\ref{eq:pf1}) leads to
    \begin{equation}
    \Delta V^{\mathcal{S}}_{{t_k}} \leqslant \left( \frac{3}{2} \kappa (x_{{t_{k+1}}},x_{{t_{k+1}}}) - \min \limits_j \kappa \left(x_{{t_{k+1}}},x_{s_j} \right) \right) \left( \Delta \alpha_{{t_k}}^{k+1} \right)^2,
    \end{equation}
    and $\Delta V^{\mathcal{S}}_{t_k} < 0$ is obtained considering (\ref{eq:thcond2}). Therefore the convergence of $\alpha^{\mathcal{S}}_{t_k}$ to $\alpha^{\mathcal{S}}_{*}$ is proven.
\end{proof}

\begin{remark}
In Theorem \ref{th:theo}, compared to (\ref{eq:thcond2}) that confines the relation between the support vectors and the new samples, (2) is determined by the distribution of future samples after $x_{t_k}$. Therefore, the sample distribution should be improved by proper control and sampling schemes such that (\ref{eq:thcond2}) and (\ref{eq:thcond1}) hold, which will be discussed in Section \ref{sec:tisd}.
\end{remark}

\subsubsection{The Sample Selection Procedure}\label{sec:tisd}
The primary reason preventing the application of incremental SVM to hybrid system identification problems is due to the heavy computational load. Due to the possible sample migrations among $\mathcal{S}$, $\mathcal{B}$ and $\mathcal{O}$, the $\alpha$ and $g$ values of every sample in $\mathcal{T}$ should be calculated in every incremental tuning. Thus the computational load drastically increases as the training set gets larger. A detailed computational analysis of incremental SVM can be found in~\cite{laskov2006incremental}.

The stability of Algorithm \ref{ag:rme} makes it possible for incremental SVM method to be applied to online identification of hybrid systems, since the number of support vectors $n_s$ used for inverse calculation is confined as $n_s = n$, and the SVM model can be trained by finite number of samples. As a result, the computation is greatly reduced compared to the original incremental SVM methods in~\cite{cauwenberghs2001incremental}. To improve the system sample distribution such that the conditions in (\ref{eq:thcond2}) and (\ref{eq:thcond2}) hold, and also further reduce the computational load, we propose the following sample selection procedure which is also mentioned by 9-11 in Algorithm \ref{ag:ptss}.

\begin{algorithm}[ht]
\caption{Procedure of Training Sample Selection}
\label{ag:ptss}
\begin{algorithmic}[1]
\STATE \textbf{procedure} tsslc($i$, $j$, $k$)
\STATE $l_k = 1$;
\STATE $l_{k-1} = -1$;
\STATE $\mathcal{T}^{ij} = \mathcal{T}^{ij} \cup x_k \cup x_{k-1}$;
\STATE \textbf{end procedure}
\end{algorithmic} 
\end{algorithm}

The general idea of Algorithm \ref{ag:ptss} is to only add two successive samples belonging to different regions to the training set when a system switching occurs. There are several advantages of this procedure,

(1). If the conditions in Assumption \ref{as:uasum} hold such that successive samples with different class labels are adequately close, then they are also more close to the switching manifold than the other samples and more likely to become support vectors. As a result, fewer samples can be used for model training while the performance is not severely affected;

(2). Since the selected samples are closely distributed around the semi-hyper-plane split by the origin $\Omega_o$, the vectorial angles between any two samples $x_{t_i}$, $x_{t_j}$ $\in \mathcal{T}$ are adequately small, such that
\begin{equation}
<x_{{t_{k+1}}},x^{\mathrm{min}}_{s_j}> \approx 0,~x^{\mathrm{min}}_{s_j} \approx \lambda x_{{t_{k+1}}},
\end{equation}
where $<\cdot,\cdot>$ denotes the angle between two vectors, $x^{\mathrm{min}}_{s_j}$ is the support vector such that $\kappa \left( x_{{t_{k+1}}},x_{s_j} \right)$ is minimum $\forall x_{s_j} \in \mathcal{S}$, and $\lambda > 1$ is a positive scalar. Therefore, $\lambda > 3/2$ leads to condition (\ref{eq:thcond2}), which holds when $x_{t_{k}}$ is adequately small under assumption \ref{as:uasum}.

(3). If the two successive samples $x_{t_r}$ and $x_{t_{r+1}}$ with different labels $l_r = - l_{r+1}$ are adequately close to each other, an approximation can be made as
\begin{equation}
\kappa \left( x_{{t_{k+1}}},x_{t_{r}} \right) \approx \kappa \left( x_{{t_{k+1}}},x_{t_{r+1}} \right),~\overline{\kappa}^{k+1}_r \approx  \overline{\kappa}^{k+1}_{i+1},~l_{k+1}l_r+ l_{k+1}l_{r+1}=0,
\end{equation}
which leads to 
\begin{equation}
l_{k+1} l_r \left( \kappa \left( x_{t_{k+1}},x_{t_r} \right) - \overline{\kappa}^{k+1}_r \right) + l_{k+1} l_{r+1} \left( \kappa \left( x_{t_{k+1}},x_{t_{r+1}} \right) - \overline{\kappa}^{k+1}_{r+1} \right) \approx 0.
\end{equation}
Since such samples are collected in pairs, it can be asserted that
\begin{equation}
\sum \limits_{i=r}^{+\infty} l_{k+1} l_r \left( \kappa \left( x_{t_{k+1}},x_{t_r} \right) - \overline{\kappa}^{k+1}_r \right) \approx 0,
\end{equation}
and condition (\ref{eq:thcond1}) is approximately satisfied.

It is worth mentioning that the conditions (\ref{eq:thcond2}) and (\ref{eq:thcond1}) do not always strictly hold, although the sample selection procedure is applied. The primary reason is due to the uncertainty when trying to approximate the original systems (\ref{eq:sys}) using finite training samples $\mathcal{T}$. The proposed methods as Algorithm \ref{ag:rme} and Algorithm \ref{ag:ptss} attempt to propose a new balance point between computational capability and estimation precision. From another perspective, the left term in (\ref{eq:thcond1}) can be recognized as an external disturbance on the incremental model training which produces perturbation on $\alpha^{S}_{t_k}$, and the sample selection procedure serves as a disturbance rejection method by improving the sample distribution.

\section{NUMERICAL SIMULATION}\label{sec:ns}
\begin{figure}[ht]
	\centering
	{\includegraphics[width=0.48\textwidth]{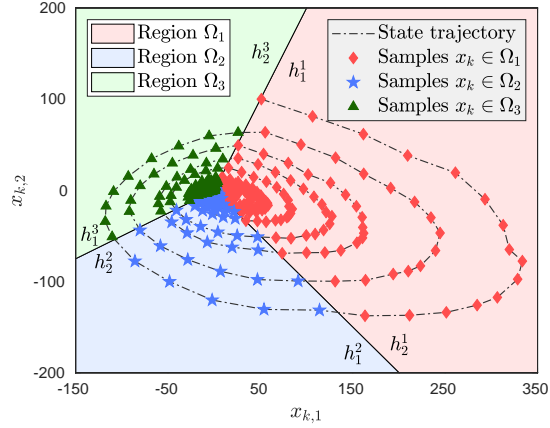}}
	\caption{The regions $\Omega^i$, switching manifolds $\chi^i$ and sequential samples $\mathcal{X}_{\tau}$ of simulated system ($i=1,2,3$). Different symbols and colors denote the belonging regions of the samples.}
	\label{fig:f1}
\end{figure}

In this section, the online identification methods for discrete time switched linear systems are evaluated by the numerical simulation of a second order PWL system with three subsystems ($r=3$). The system state and input vectors are represented as $x_{t_k} = \left[x^1_{t_k},~x^2_{t_k} \right]^T \in \mathbb{R}^2$ and $u_{t_k} \in \mathbb{R}$, the parameters of the subsystems are respectively
\begin{equation}\label{eq:sysmat}
\begin{split}
A^1 = \left[ \begin{array}{cc}
0.9 & 0.3 \\ -0.2 & 0.75
\end{array} \right],~A^2 &= \left[ \begin{array}{cc}
0.8 & 0.25 \\ -0.3 & 0.7
\end{array} \right],~A^3 = \left[ \begin{array}{cc}
0.7 & 0.4 \\ -0.35 & 0.9
\end{array} \right], \\
B^1 = \left[ \begin{array}{cc}
1.5 & 0.8
\end{array} \right]^T,~&B^2 = \left[ \begin{array}{cc}
1 & 1
\end{array} \right]^T,~B^3 = \left[ \begin{array}{cc}
0.9 &  1.2
\end{array} \right]^T,
\end{split}
\end{equation}
and the sets of switching manifolds $\chi^1$, $\chi^2$ and $\chi^3$ respectively surrounding the regions $\Omega^1$, $\Omega^2$ and $\Omega^3$ are
\begin{equation}
\begin{split}
&\chi^1:\chi^1_1 = \left\{ x \left| x^T h^1_1,~x_1 > 0 \right. \right\},~\chi^1_2 = \left\{ x \left| x^T h^1_1,~x_1 > 0 \right. \right\}, \\
&\chi^2:\chi^2_1 = \left\{ x \left| x^T h^2_1,~ -x_1 > 0 \right. \right\},~\chi^2_2 = \left\{ x \left| x^T h^2_1,~x_1 > 0 \right. \right\}, \\
&\chi^3:\chi^3_1 = \left\{ x \left| x^T h^3_1,~ -x_1 > 0 \right. \right\},~\chi^3_2 = \left\{ x \left| x^T h^3_1,~-x_1 > 0 \right. \right\},
\end{split}
\end{equation}
where each manifold is respectively represented as
\begin{equation}\label{eq:sysman}
\begin{split}
h^1_1 = \left[ \begin{array}{cc}
-2 & 1
\end{array} \right]^T,~h^2_1 = \left[ \begin{array}{cc}
1 & 1
\end{array} \right]^T,~h^3_1 = \left[ \begin{array}{cc}
1 & -2
\end{array} \right]^T,\\ h^3_2 = \left[ \begin{array}{cc}
2 & -1
\end{array} \right]^T,~h^1_2 = \left[ \begin{array}{cc}
-1 & -1
\end{array} \right]^T,~h^2_2 = \left[ \begin{array}{cc}
-1 & 2
\end{array} \right]^T,
\end{split}
\end{equation}
and we use the following vector sets $\mathcal{H}^1$, $\mathcal{H}^2$ and $\mathcal{H}^3$ to denote the manifolds $\chi^1$, $\chi^2$ and $\chi^3$,
\begin{equation}
\mathcal{H}^1 = \left\{h^1_1,~h^1_2 \right\},~\mathcal{H}^2 = \left\{ h^2_1,~h^2_2 \right\},~\mathcal{H}^3 = \left\{ h^3_1,~h^3_2 \right\}.
\end{equation}
It has been noticed that $h^1_1 = - h^3_2$, $h^2_1 = - h^1_2$ and $h^3_1 = - h^2_2$, which represents that $\chi^1_1$ and $\chi^3_2$, $\chi^2_1$ and $h^1_2$, $\chi^3_1$ and $h^2_2$ are respectively overlapped with each other. The initial system state is defined as $x_0 = \left[~51~~100~\right]^T \in \Omega^1$, and the control law $u_{t_k} = \left(L + \varepsilon_{t_k} \right) x_k$ is applied to the system, where $L = \left[~0.12~~0.12~\right]$ is the state feedback control gain given without further interpretations, and $\varepsilon_k$ is a Gaussian noise sequence following the distribution $\varepsilon_k \sim N(0,0.005)$ added to improve the sample distribution. The system runs for $\tau = 3600$ samples, and the input and state sequences $\mathcal{U}_{\tau}$ and $\mathcal{X}_{\tau}$ are produced. The regions $\Omega_1$, $\Omega_2$, $\Omega_3$, switching manifolds $\mathcal{H}_1$, $\mathcal{H}_2$, $\mathcal{H}_3$ and sequential state samples $\mathcal{X}_{\tau}$ are illustrated in Fig. \ref{fig:f1}, and it is not difficult to infer that $\mathcal{X}_{\tau}$ satisfies all conditions in Assumption \ref{as:uasum}.

The goal of the simulation is to estimate the values of the system parametric matrices in (\ref{eq:sysmat}) and the switching manifold vectors in (\ref{eq:sysman}) based on the input and state sample sequences $\mathcal{U}_{\tau}$ and $\mathcal{X}_{\tau}$. We simulate the sequential sampling along the state trajectory (as shown in Fig. \ref{fig:f1}) and simultaneously run the parameter estimator (\ref{eq:law}), Algorithm \ref{ag:dtc} and Algorithm \ref{ag:rme} with the training samples selection and the support vector pruning procedures. The parameter $\Gamma$ in (\ref{eq:law}) is selected as $\Gamma = 0.9 I_2$ where $I_2$ is the two dimensional identity matrix, the matrix $P$ for the kernel function (\ref{eq:kernel}) is simply determined as $P = I_2$.

After the simulating the $\tau$ sequential samples, the results show that the final estimated parameters $\hat{A}^1_{\tau}$, $\hat{A}^2_{\tau}$, $\hat{A}^3_{\tau}$, and $\hat{B}^1_{\tau}$, $\hat{B}^2_{\tau}$, $\hat{B}^3_{\tau}$ are of the same value as the true parameter values in (\ref{eq:sysmat}), and 3595 out of 3600 samples are labeled with the correct region numbers, marking a successful rate of system switching detection as $99.86\%$. Thus the precision and accuracy of the parameter estimation and switching detection are confirmed. Additionally, the final weight values of the SVM models are
\begin{equation}\label{eq:estman}
w^{12} = \left[ \begin{array}{r}
-19.8674 \\ -20.4493
\end{array} \right]^T,~
w^{23} = \left[ \begin{array}{r}
-14.0526 \\ 29.3731
\end{array} \right]^T,~
w^{31} = \left[ \begin{array}{r}
27.4342 \\ -13.4117
\end{array} \right]^T.
\end{equation}
Since the manifolds are assumed to pass the origin $\Omega_o$, we compare the estimated results (\ref{eq:estman}) and the true values of the switching manifolds (\ref{eq:sysman}) by calculating their vectorial slopes, i.e. the ratio between the second and the first components. Therefore, the manifold estimation error can be defined as
\begin{equation}
\left| \tilde{S}^{ij}_k \right| = \left| S\left(w^{ij} \right)- S\left(h^i_{1} \right) \right| = \left| \frac{\left(w^{ij}\right)_2}{\left(w^{ij}\right)_1} - \frac{\left(h^{i}_{1}\right)_2}{\left(h^{i}_{1}\right)_1} \right|,~i,j=1,2,3.
\end{equation}
Note that the singular cases are ignored here. After the calculation, we obtain $\tilde{S}^{12}_{\tau} = 0.0293,~\tilde{S}^{12}_{\tau} = 0.0902,~\tilde{S}^{12}_{\tau} = 0.0111$, which confirms the estimation precision of the  switching manifolds.

\begin{figure}[htpb]
	\centering
	\subfigure[The recursive train process of the SVM models.]{\label{fig:f21}\includegraphics[width=0.48\textwidth]{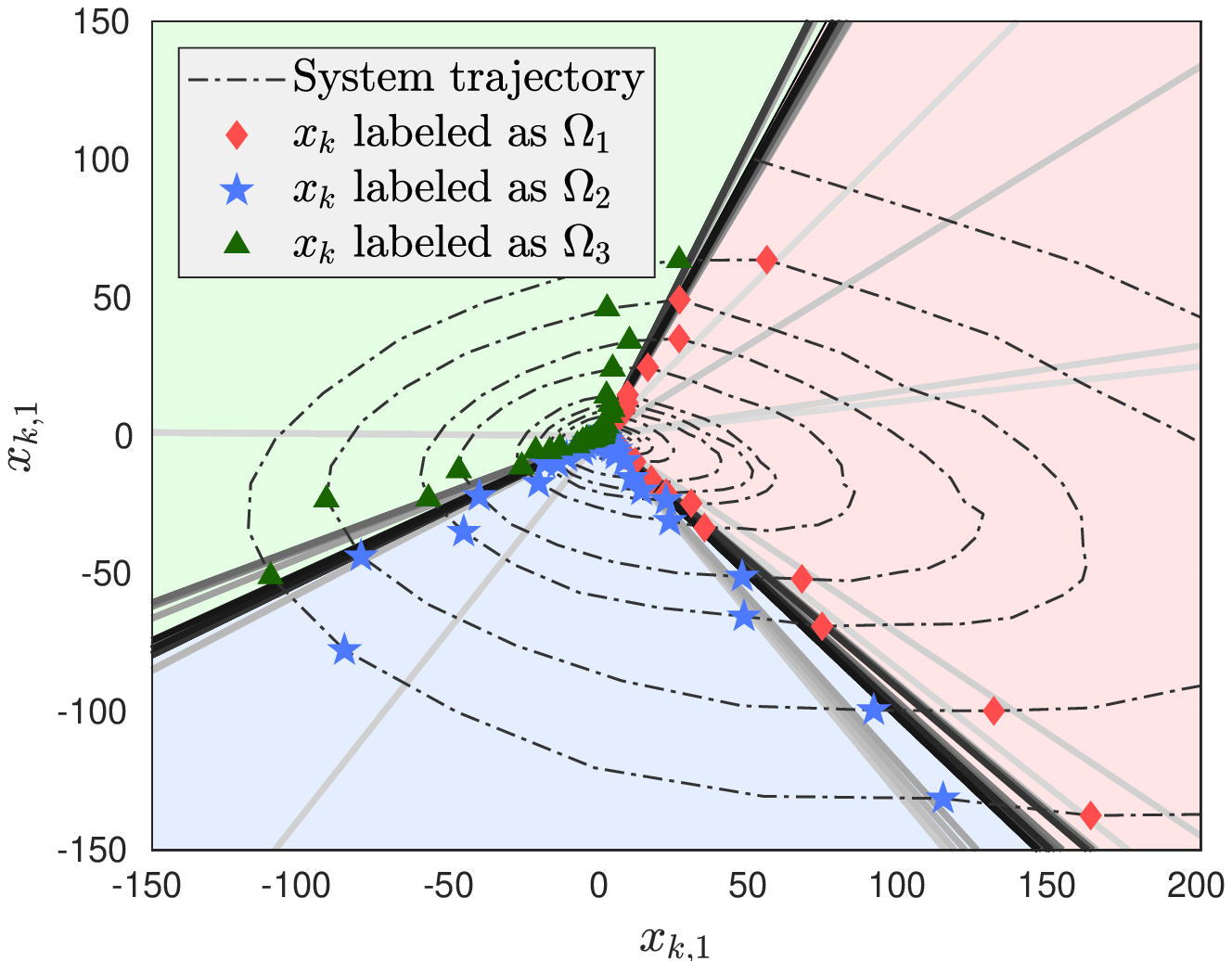}}		
	\subfigure[The finally trained SVM models and estimated manifolds.]{\label{fig:f22}\includegraphics[width=0.48\textwidth]{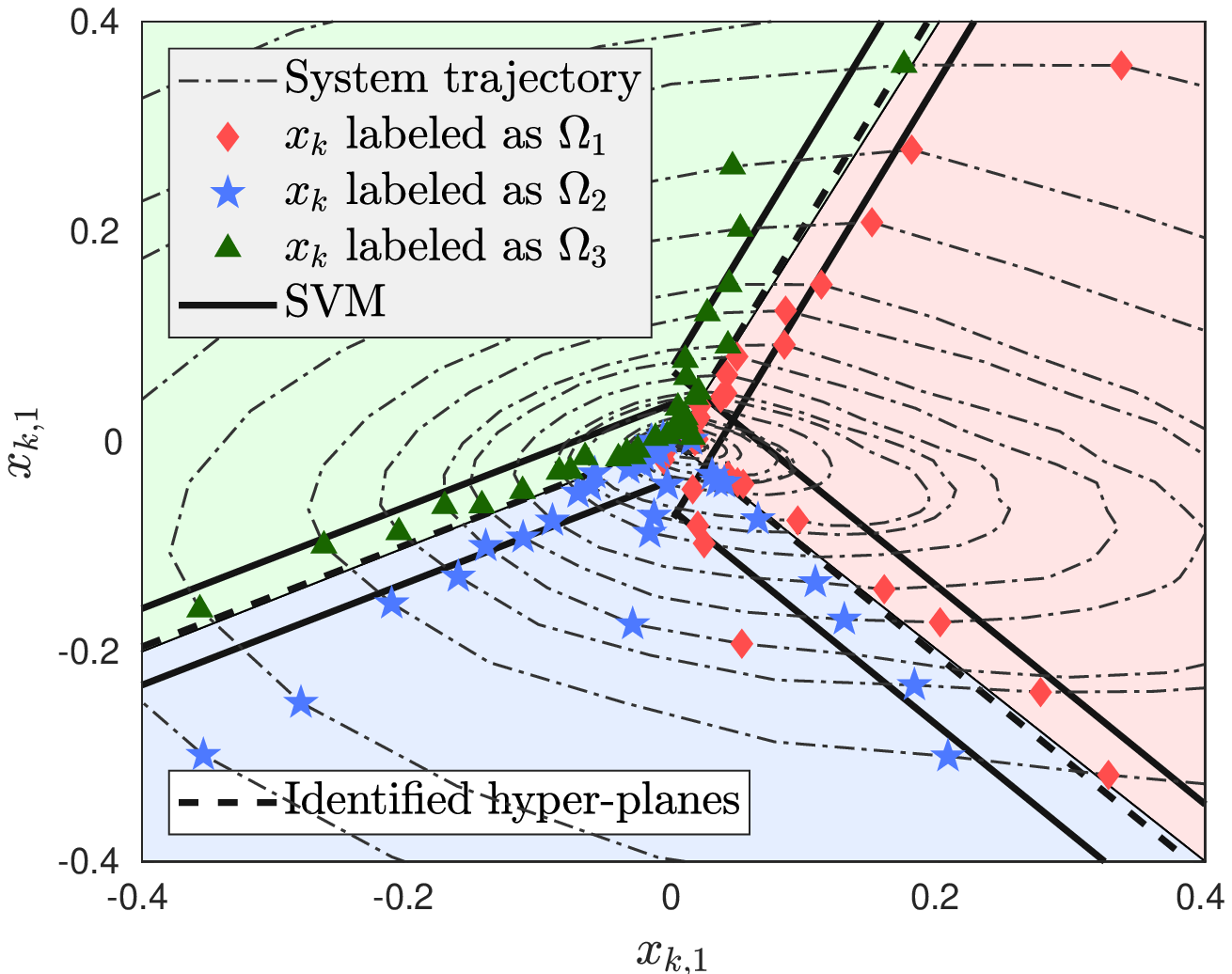}}		
	\caption{The selected training samples $\mathcal{T}$ and the recursive estimation process of the switching manifolds. Different symbols and colors denote the estimated labels of the samples.}
	\label{fig:f2}
\end{figure}

\begin{figure}[htpb]
	\centering
	\subfigure[Estimation errors $\left\| \tilde{\Phi}^i_k \right\|_2$.]{\label{fig:f3}\includegraphics[width=0.48\textwidth]{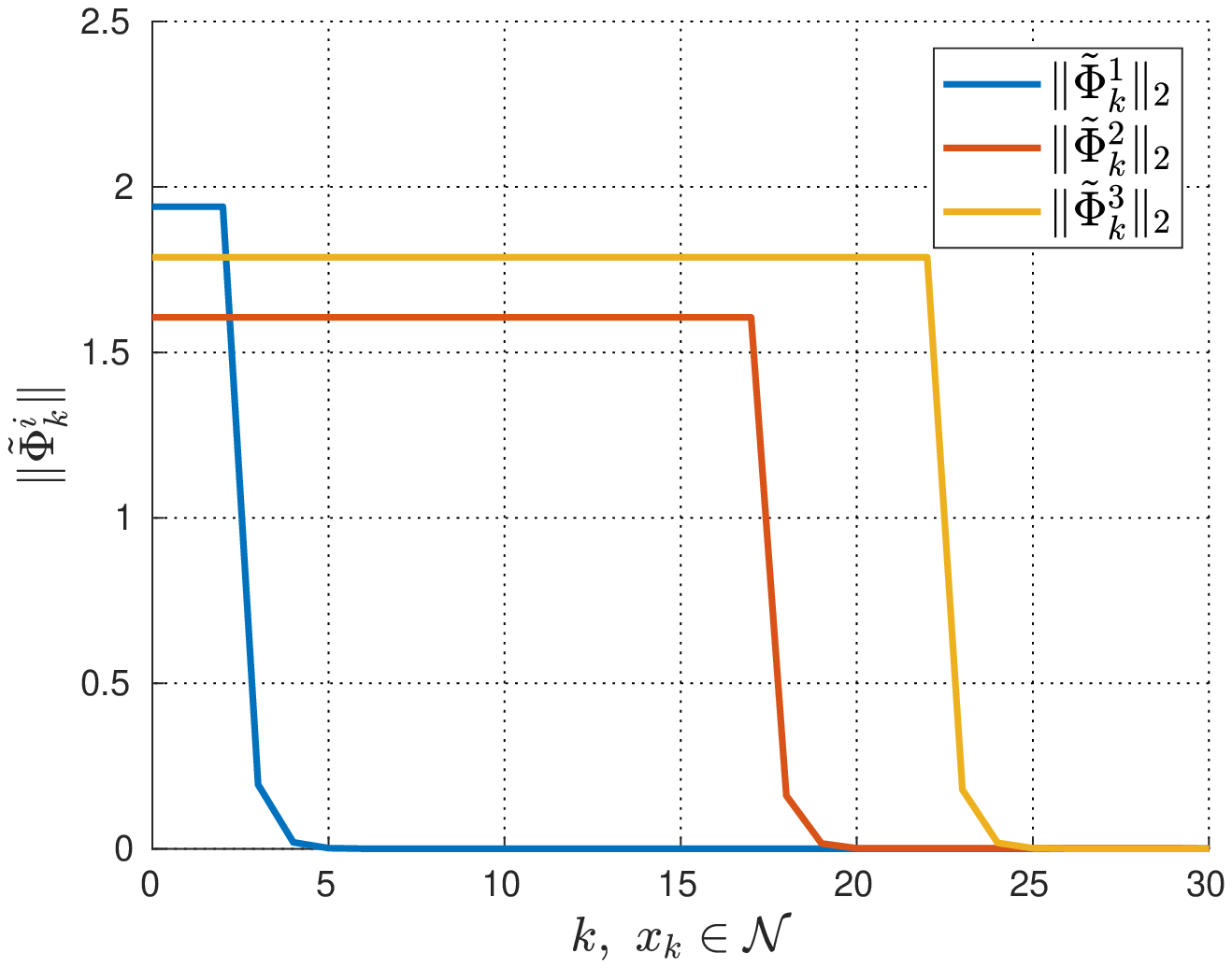}}		
	\subfigure[Estimation errors $\tilde{S}^{ij}_k$.]{\label{fig:f4}\includegraphics[width=0.48\textwidth]{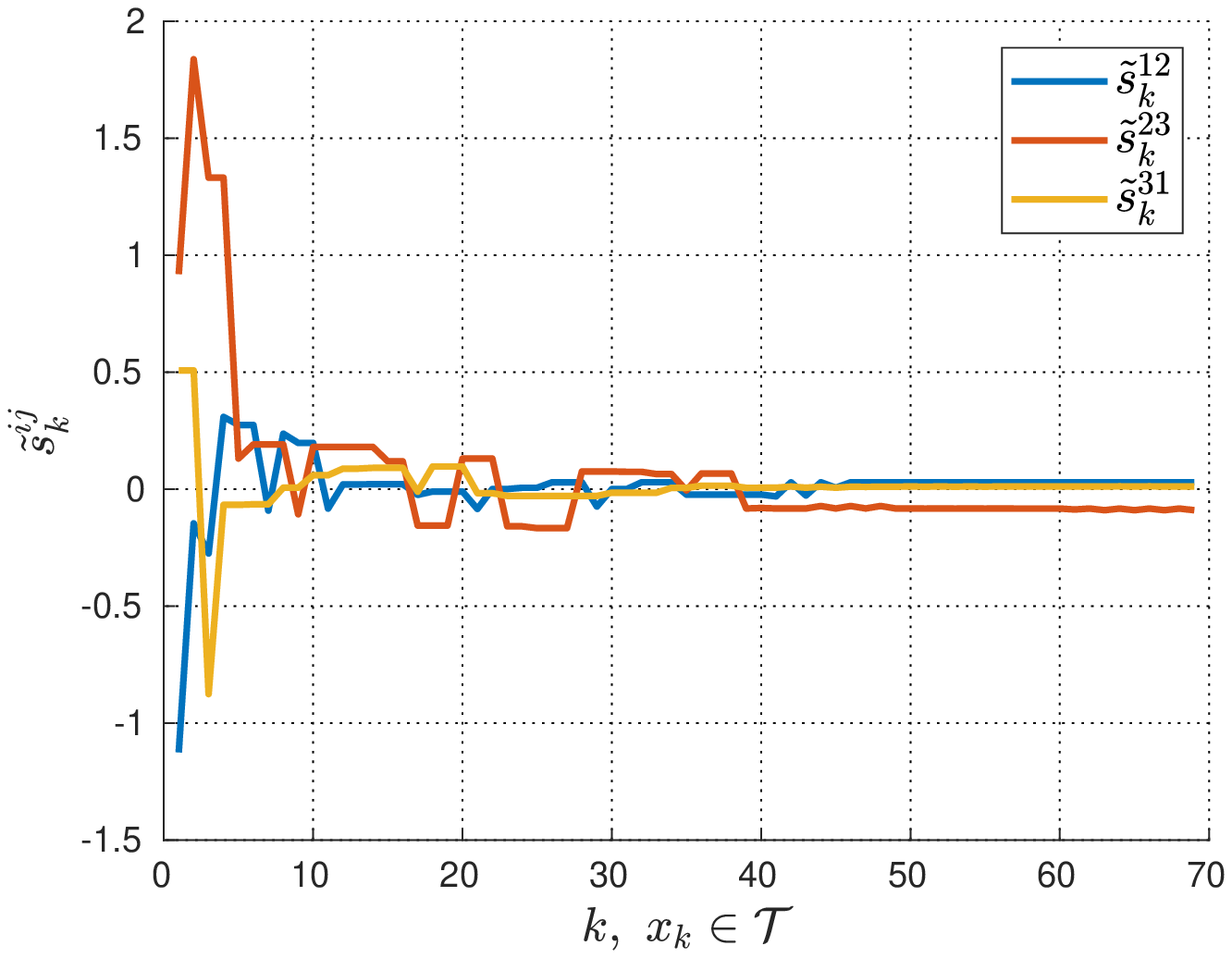}}		
	\caption{The convergence of parameter estimation errors.}
	\label{fig:f34}
\end{figure}



The simulation results are also illustrated in the following figures. The selected training samples and the incremental training processes for recursive switching manifold estimation are shown in Fig. \ref{fig:f21}, and the finally trained SVMs and estimated switching manifolds are illustrated in \ref{fig:f22} with a larger zoomed scale. Meanwhile, the convergence of estimation errors of subsystem parameters $\left\| \tilde{\Phi}^i_k \right\|_2$ and switching manifolds $\left\| \tilde{w}^{ij}_k \right\|_2$ as $k$ increases are respectively shown in Fig. \ref{fig:f3} and Fig. \ref{fig:f4}. 

The Fig. \ref{fig:f2} indicates that only the samples closed to the switching manifolds are selected as the training sets. The incrementally tuned svm models in each time when new samples are obtained are shown in Fig. \ref{fig:f21}, where the changing of the models are depicted by the gradually varied colors, and the converging tendency of the models to the true switching manifolds is obvious. A clear illustration of the final SVM models and their corresponding estimated manifolds is shown in Fig. \ref{fig:f22}. It is seen that the estimated manifolds are quite close to the true manifolds, which confirms the precision of switching manifold estimation.

Meanwhile, the convergence of estimation errors $\left\|\tilde{\Phi} \right\|$ and $\tilde{S}^{ij}_k$ as $k$ increases are shown in Fig. \ref{fig:f3} and Fig. \ref{fig:f4}. It is noticed in Fig. \ref{fig:f3} that the estimation $\hat{\Phi}^1_k$, $\hat{\Phi}^2_k$ and $\hat{\Phi}^3_k$ quickly converge to the true values $\Phi^1$, $\Phi^2$ and $\Phi^3$ at the sampling instants respectively $k = 5$, $k=20$ and $k=25$. Note that the platforms on the figure lines before these instants reveal the initialization of the history stacks $R^i_k$ and $D^i_k$, while the descending lines represent the real converging times. In Fig. \ref{fig:f4}, fluctuations can be observed in the earlier stages before $k = 40$, while after this instant, the estimation error $\tilde{S}^{ij}_k$ converges close to zero, although small deviations caused by distribution uncertainties may exist (eg. $\tilde{S}^{23}_k$).

\section{conclusion}\label{sec:cf}
In this paper, we propose a novel online identification framework for a class of discrete time switched linear systems. Specifically, the problems of recursive parameter estimation, online system switching detection and recursive switching manifold estimation are solved by applying techniques of concurrent learning and incremental SVM. Taking use of the history stacks collecting historical system samples, the exponential convergence of parameter estimation is guaranteed and sample labeling is correctly conducted with non-zero estimation errors. Due to the support vector tuning and sample selection procedures, the classification SVM model can be trained using finite samples, and the stability of the switching manifold estimation algorithm is also guaranteed. As a pure online identification procedure, the estimation precision and light computation are confirmed by theoretical proofs and simulation results.

Nevertheless there are still several issues remaining unstudied in this paper which should be specifically investigated in the future. First of all, further studies on how to design the input control sequences that satisfy the conditions in Assumption \ref{as:uasum} are still necessary. The adaptive control scheme is a potential method, since it is capable of producing the control sequence and estimating parameters at the same time. Moreover, the proposed online identification framework is to be generalized for a broader class of hybrid systems with less conservative assumptions. Finally, the relation between the system input and sample distribution is also an interesting topic to be investigated, such that stability and precision of the switching manifold estimation are further improved.

\bibliographystyle{unsrt}  
\bibliography{references}

\end{document}